\documentclass[aps,prb,twocolumn,amsmath,showpacs,showkeys,floatfix]{revtex4}
\usepackage{amsmath}
\usepackage{amssymb}
\usepackage{latexsym}
\usepackage{epsfig}
\usepackage{subfigure}
\usepackage{xspace}
\usepackage{color}

\newcommand{\vc}[1]{\mathbf{#1}}

\newcommand{\ozz}{$\text{O}_\text{z}^Z$\xspace}
\newcommand{\obx}{$\text{O}_\text{B}^X$\xspace}

\newcommand{\dhalbe}{$\Delta_1/2$\xspace}

\begin{document}
\thispagestyle{empty}

\title{A microscopic modeling of phonon dynamics and charge response in NdCuO}
\author{Thomas Bauer}
\author{Claus Falter}
\email[Email to: ]{falter@nwz.uni-muenster.de}
\affiliation{Institut f\"ur Festk\"orpertheorie, Westf\"alische
Wilhelms-Universit\"at,\\ Wilhelm-Klemm-Str.~10, 48149 M\"unster,
Germany}
\date{\today}

\pacs{74.72.Jt, 74.25.Kc, 71.38.-k, 63.20.Dj}

\keywords{high-temperature superconductors, lattice dynamics,
electronic density response, electron-phonon interaction}

\begin{abstract}
A description of phonon dynamics and charge response of the $n$-doped
high-temperature superconductor (HTSC) NdCuO is presented based upon a
microscopic  modeling of the electronic density response. This is
accomplished starting from the insulating state via the underdoped
strange metallic to the more conventional metallic state by consecutive
orbital selective incompressibility-compressibility transitions in
terms of strict sum rules for the charge response. The approach
proposed in this work for the $n$-doped HTSC's modifies the modeling
recently applied to the $p$-doped compounds and expresses an
electron-hole asymmetry introduced by doping. A qualitative physical
picture consistent with our modeling of the electronic state in the
cuprates is given in which a sufficiently broad set of orbital degrees
of freedom, i.e. Cu$3d$/$4s$ and O$2p$ is essential. Within the
framework of linear response theory we calculate full phonon dispersion
curves in the different phases. In particular, the strongly doping
dependent anomalous high-frequency oxygen bond-stretching modes (OBSM)
found experimentally in the $p$-doped HTSC's and also recently for
$n$-doped Nd$_{1.85}$Ce$_{0.15}$CuO$_{4}$ are investigated and compared
with experimental results from inelastic neutron scattering and
inelastic x-ray scattering, respectively. We calculate an anticrossing
scenario for the OBSM in $n$-doped NdCuO which is absent in the case of
$p$-doped LaCuO and relate it to the different crystal structure.
Phonon-induced electronic charge redistributions of the anomalous OBSM
due to nonlocal electron-phonon interaction effects of
charge-fluctuation type giving reason to dynamic stripes are also
studied. Finally, calculations of a characteristic rearrangement of the
phonon density of states across the insulator-metal transition are
presented and a comparison with experimental results has been
accomplished.
\end{abstract}

\maketitle
\section{INTRODUCTION}\label{sec_intr}
There is increasing evidence that the electron-phonon coupling is
strong in the cuprate based high-temperature superconductors (HTSC's)
and phonons might play an important role for the electron dynamics in
the HTSC's, see e.g. Refs. \onlinecite{1,2,3,4,5,6,7,8}. It has been
known from the experiments for some time that the frequencies of the
high-frequency oxygen bond-stretching modes (OBSM) are strongly
renormalized (softened) upon doping in the $p$-doped cuprates
investigated so far, Refs. \onlinecite{3,9,10,11,12,13,14,15}. More
recently, an anomalous softening behaviour of the OBSM, not present in
the undoped insulating state, also has been observed in the $n$-doped
metallic state of NdCuO, Refs. \onlinecite{16,17,18}. These findings
support the generic nature of the anomalies in all HTSC's. The
frequency renormalization and a corresponding increase of the phonon
linewidths point to a strong coupling of these phonons to the charge
carriers. Theoretical calculations for the $p$-doped cuprates have
shown that the strong \textit{nonlocal} electron-phonon interaction
(EPI) leading to the softening of the OBSM is due to a specific
screening effect in terms of a \textit{local} polarization created by
metallic charge fluctuations (CF's) on the outer shells of the Cu an
O$_\text{xy}$ ions in the CuO-plane, see e.g. Refs.
\onlinecite{19,20,21}.

This type of screening in a strongly inhomogeneous electronic system
induced by a density distribution which mainly is concentrated at the
ions in the CuO-plane leads for the OBSM in the $p$-doped
cuprates\cite{19,21} and also for $n$-doped NdCuO, as shown in this
work, to an anomalous behaviour of the dispersion. The specific
screening yields a \textit{downward} dispersion of the longitudinal
branch along the (1,0,0) and the (1,1,0) direction falling in the
(1,0,0) direction in LaCuO and partly also in NdCuO even below the
transverse branch. Such a behaviour cannot be understood within a
typical lattice dynamical model, like e.g. the shell model, including a
homogeneous electrongas screening that is \textit{diagonal} in
reciprocal space. The latter type of dielectric response always gives
an \textit{increasing} dispersion when passing from the center of the
Brillouin zone (BZ) into the zone, because of incomplete screening of
the changes of the Coulomb potential at shorter distance. Such a
behaviour is found in our calculations for the dispersion of the
branches for modes with polar character at lower energy.

Contrary, the anomalous softening of the OBSM leading to a decreasing
dispersion is a result of an ''over-screening'' at shorter wave lengths
of the changes of the Coulomb potential generated by the motion of the
ions in these particular modes. Physically, this ''over-screening'' in
the OBSM giving rise to bond length modulations has been shown by our
calculations to be due to nonlocally excited ionic CF's localized at
the Cu and O$_\text{xy}$ sites leading to a (dynamic) charge ordering
in form of localized stripes of alternating sign in the CuO plane.
Expressed in terms of the density response matrix in reciprocal space
such strong nonlocal EPI effects and ''over-screening'' of the OBSM is
related to their \textit{off-diagonal} elements (local-field effect)
being apparently very important in the cuprates as revealed by the
anomalous softening.

It is interesting to note in this context, that an ''over-screening''
effect is not found in the in-plane OBSM of metallic SrRuO, a
low-temperature, T$_c$=1.5K, (presumably) triplet superconductor,
structurally isomorphic to LaCuO, see Ref. \onlinecite{22}. This points
to a more homogeneous electrongas-like screening mechanism in this
material and correspondingly to a reduction of the strong nonlocal EPI
found in the cuprates. The different screening in SrRuO as compared
with the cuprates should be associated with a more extended Ru$4d$
state as compared to the localized, stronger correlated Cu$3d$ state
and a strong hybridization between Ru$4d$ and O$2p$. Furthermore, SrRuO
is not a low-carrier metal and has a large density of states at the
Fermi energy.

In our microscopic theoretical description of the charge response, the
EPI and lattice dynamics of the HTSC's in the framework of the linear
response approach we focus on the specific features of their
solid-state chemistry which is necessary for a realistic description of
the cuprates since the phenomena observed in the HTSC's hardly occur in
the other known compounds.

The strong ionic nature of the cuprates is modeled in our approach by
an \textit{ab initio} rigid ion model (RIM), used as a reference frame
for the local ionic, rigid charge response and the EPI, respectively.
The important and characteristic nonlocal, nonrigid contribution to the
electronic density response and the EPI in the HTSC's is calculated in
terms of microscopically well defined CF's and dipole fluctuations
(DF's). The CF's and DF's are excitable on the electronic shells of the
ions in the crystal. Additionally, covalent metallic features of
bonding are approximatively taken into account \cite{19,23}. In the
calculations a sufficiently broad set of orbital degrees of freedom
(Cu$3d$, Cu$4s$ and O$2p$), the full three-dimensional long-ranged
Coulomb interaction as well as short-ranged local repulsions of the
electrons, in particular the important on-site repulsion mediated by
the localized Cu$3d$ orbitals, is considered quantitatively. All the
resulting couplings arising in the dynamical matrix and the EPI are
microscopically well defined and can be calculated. The results
obtained for the $p$-doped materials so far agree well with the
experimental phonon dispersion. For a recent review, see Ref.
\onlinecite{19}.

Our model for the (irreducible) electronic polarizability matrix,
$\Pi_{\kappa\kappa'}$, specifying the kinetic part of the charge
response\cite{19,24}, is defined in a localized basis by the orbital
degrees of freedom $\kappa,\kappa'$ in the elementary cell of the
crystal. In the calculations for the HTSC's Cu$3d$, Cu$4s$ and O$2p$
orbitals in the CuO plane are taken into account.

In the absence of a rigorous quantitative description of the electronic
state of the HTSC's our model for $\Pi_{\kappa\kappa'}$ has been
designed for the different phases to be consistent with the
corresponding charge response in the long-wavelength limit. This is
achieved by fulfilling rigorous sum rules for the electronic density
response in terms of the polarizability matrix \cite{20,23}. These sum
rules, see also Sec. \ref{sec_theory}, can be considered as orbital
resolved closed forms to represent the change of the charge response
across a metal-insulator transition in terms of the electronic
polarizability or the compressibility of the electronic system,
respectively, as a primary tool to characterize the corresponding
ground state.

Hitherto such a modeling of the electronic state of the cuprates only
has been applied to the $p$-doped materials, see Refs.
\onlinecite{19,21,23} for a detailed discussion. We start from the
insulating state of the HTSC's where both the Cu$3d$ and O$2p$ orbitals
are taken to be incompressible, consistent with a gap in the charge
excitation spectrum, and the delocalized Cu$4s$ orbitals are neglected.

For the underdoped "strange metallic" phase of the $p$-doped HTSC's,
where the description of the low-energy excitations remains a
theoretical challenge, we have proposed a pseudogap model for the
electronic density response in the past \cite{23}. It describes the
localization-delocalization transition in the electronic structure in
terms of a qualitatively different compressibility of the Cu$3d$ and
O$2p$ states, respectively, and admits that the single-particle density
of states is suppressed at the Fermi energy for the strongly localized
and correlated Cu$3d$ states because of the cost in energy from hopping
of the charge carriers to the Cu-sites. This is achieved by installing
an insulator-like, incompressible charge response  for the localized
Cu$3d$ states via the sum rules, but not so for the more delocalized
O$2p$ orbitals where the holes predominantly are injected to in
$p$-type cuprates. For the O$2p$ orbitals a compressible, metallic
charge response is allowed with a renormalized partial density of
states at the Fermi energy $\varepsilon_{F}$ consistent with a gain in
kinetic energy by delocalization.

The blocking of the metallic charge response at the Cu sites in the
underdoped phase is lifted in our modeling of the optimally and
overdoped metallic phase by allowing all orbitals, i.e. Cu$3d$, Cu$4s$
and O$2p$, to become compressible, metallic with a nonvanishing partial
density of states at $\varepsilon_{F}$ for \textit{all} orbitals
considered. In a $k$-space picture a large Fermi surface (FS) as found,
e.g., in LDA-calculations, can now be expected to develop from a gapped
FS of the underdoped phase seen in the experiments by angle-resolved
photoemsission (ARPES).

Thus, according to the modeling, we have a crossover in the electronic
properties corresponding to a qualitative change of the ground state to
a Fermi-liquid like state in the overdoped material. Apparently, the
behaviour is quite different on the two sides of the crossover point,
and there is no phase transition with symmetry breaking assumed in our
modeling.

For the $n$-doped  materials we investigate quantitatively in this work
a model proposed recently in Ref. \onlinecite{19} that modifies the
sequence of the orbital selective incompressibility-compressibility
transitions. In this way an \textit{electron-hole asymmetry} is
introduced in our modeling of the insulator-metal transition,
simulating the different character of hole and electron doping. While
in the $p$-doped HTSC's hole carriers are brought to the O$2p$ orbitals
making the latter compressible in our picture, in the $n$-doped
materials, like Nd$_{2 - x}$Ce$_{x}$CuO$_{4}$, electron carriers are
brought into orbitals with Cu$3d$-4$s$ character at the Cu sites. Thus,
we model with the help of the sum rules an underdoped $n$-type material
in our approach in terms of a metallic, compressible charge response at
the Cu sites (Cu$3d$) with an increasing  Cu$4s$ component upon doping
and a localized incompressible, insulator-like response of the O$2p$
orbitals at the O$_{xy}$ sublattices of the CuO plane.

Such a modeling also may be consistent with a stabilization of
antiferromagnetic spin correlations, because an insulator-like
localization of the incompressible O$2p$ orbitals should be favourable
for superexchange. Experimentally it is found that the phase diagram is
asymmetric with respect to electron and hole doping, and for the
$n$-type materials the antiferromagnetic phase extends much further
with doping. From our results for the electronic CF polarizability of
the oxygen one might speculate that this quantity should play an
indirect role for antiferromagnetism and superconductivity in the
cuprates. Moreover, the incompressible response for O$2p$ also
generates a pseudogap phenomenon attenuating the density of states at
$\varepsilon_{F}$.

Then, in the optimally doped state a crossover to a metallic charge
response at the formerly incompressible O$2p$ orbitals is admitted and
so \textit{all} orbitals, i.e. Cu$3d$, Cu$4s$ and O$2p$ are modeled as
compressible by allowing the O$2p$ states to become compressible too,
like in the $p$-doped case. Correspondingly a change of the FS topology
can be expected and Fermi-liquid behaviour should revive in the system.

Note in this context that from ARPES experiments, Refs.
\onlinecite{25,26}, of the $n$-type superconductor Nd$_{2 -
x}$Ce$_{x}$CuO$_{4}$ it is found, that at low doping the FS is a small
electron pocket (with volume $\sim x$) centered at $(\pi,0,0)$. Further
doping finally leads to a large LDA-like FS of the hole-type (volume
$\sim 1 + x$) centered at $(\pi,\pi,0)$. A change of FS topology upon
doping, different from the $n$-doped material, also has been observed
in the $p$-doped HTSC's. Here, at low doping ARPES experiments reveal
so-called "Fermi surface arcs" (which also could be elongated Fermi
surface pockets at nodal positions). With further doping in the
optimally doped and overdoped state a large LDA-like FS of the
hole-type forms as in the $n$-doped case, Refs. \onlinecite{27,28}.

The article is organized as follows. In Sec. \ref{sec_theory} elements
of the theory and modeling are reviewed to set the frame. Section
\ref{sec_dispersions} presents our calculated results of the phonon
dispersion of NdCuO across the insulator-metal transition taking CF's
and DF's in the electronic density response into account. In Sec.
\ref{sec_anticrossing} the possibly generic phonon anomalies and the
anticrossing behaviour is discussed, while Sec.
\ref{sec_chargeresponse} deals with the phonon-induced charge response
of the OBSM. In Sec. \ref{sec_dos} the phonon density of states for
insulating and metallic NdCuO is calculated and a comparison of the
redistribution of spectral weight is provided. Finally, a summary of
the paper is presented in Sec. \ref{sec_summary} and the conclusions
are given.

\section{THEORY AND MODELING}\label{sec_theory}
From a general point of view our treatment of the electronic density
response and lattice dynamics in terms of DF's and CF's can be
considered as a microscopic implementation of the phenomenological
dipole-shell model or the charge-fluctuation models, respectively. For
a general formulation of phenomenological models for lattice dynamics
that use localized electronic variables as adiabatic degrees of
freedom, see for example Ref. \onlinecite{29}. This formulation covers
shell models, bond-charge models and charge-fluctuation models. While
in this approach the coupling coefficients are treated as empirical
fitting parameters, the essential point in our microscopic scheme is
that all the couplings can be calculated.

In the following a survey of the theory and modeling is presented. A
detailed description can be found in Ref. \onlinecite{20} and in
particular in Ref. \onlinecite{30} where the calculation of the
coupling parameters is presented.

The local part of the electronic charge response and the EPI is
approximated in the spirit of the \textit{Quasi-Ion concept} \cite{31}
by an ab initio rigid ion model (RIM) taking into account covalent ion
softening in terms of (static) effective ionic charges calculated from
a tight-binding analysis (TBA). In addition, scaling of the
short-ranged part of certain pair potentials between the ions is
performed to simulate further covalence effects in the calculation in
such a way that the energy-minimized structure is as close as possible
to the experimental one \cite{32}. Structure optimization and energy
minimization is very important for a reliable calculation of the phonon
dynamics through the dynamical matrix.

The RIM with the corrections just mentioned then serves as an unbiased
reference system for the description of the HTSC's and can be
considered as a first approximation for the insulating state of these
compounds.  Starting with such an unprejudiced rigid reference system
nonlocal, nonrigid electronic polarization processes are introduced in
form of more or less localized electronic charge-fluctuations (CF's) at
the outer shells of the ions. Especially in the metallic state of the
HTSC's the latter dominate the nonlocal contribution of the electronic
density response and the EPI and are particularly important in the CuO
planes. In addition, \textit{anisotropic} dipole-fluctuations (DF's)
are admitted in our approach, Refs. \onlinecite{24,30}, which prove to
be specificly of importance for the ions in the ionic layers mediating
the dielectric coupling and for the polar modes. Thus, the basic
variable of our model is the ionic density which is given in the
perturbed state by
\begin{equation}\label{1}
\rho_\alpha(\vc{r},Q_\lambda, \vc{p}_\alpha) =
\rho_\alpha^0(r) + \sum_{\lambda}Q_\lambda \rho_\lambda^\text{CF}(r)
+ \vc{p}_\alpha \cdot
\hat{\vc{r}} \rho_\alpha^\text{D}(r).
\end{equation}
$\rho_{\alpha}^{0}$ is the density of the unperturbed ion, as used in
the RIM, localized at the sublattice $\alpha$ of the crystal and moving
rigidly with the latter under displacement. The $Q_{\lambda}$ and
$\rho^{\text{CF}}_{\lambda}$ describe the amplitudes and the form
factors of the CF's and the last term in Eq. \eqref{1} represents the
dipolar deformation of an ion $\alpha$ with amplitude (dipole moment)
$\vc{p}_{\alpha}$ and a radial density distribution
$\rho_{\alpha}^{\text{D}}$ as form factor.  $\hat{\vc{r}}$ denotes the
unit vector in the direction of $\vc{r}$. The
$\rho^{\text{CF}}_{\lambda}$ are approximated by a spherical average of
the orbital densities of the ionic shells calculated in LDA taking
self-interaction effects (SIC) into account. The dipole density
$\rho_{\alpha}^{\text{D}}$ is obtained from a modified Sternheimer
method in the framework of LDA-SIC, Ref. \onlinecite{30}. All
SIC-calculations are performed for the average spherical shell in the
orbital-averaged form according to Perdew and Zunger, Ref.
\onlinecite{33}. For the correlation part of the energy per electron
$\epsilon$ the parametrization given in Ref. \onlinecite{33} has been
used.

The total energy of the crystal is obtained by assuming that the
density can be approximated by a superposition of overlapping densities
$\rho_{\alpha}$. The $\rho_{\alpha}^{0}$ in Eq. \eqref{1} are also
calculated within LDA-SIC taking environments effects, via a Watson
sphere potential and the calculated static effective charges of the
ions into account. The Watson sphere method is only used for the oxygen
ions and the depth of the Watson sphere potential is set as the
Madelung potential at the corresponding site. Such an approximation
holds well in the HTSC's \cite{32,34}. As a general rule, partial
covalence reduces the amplitude of the static effective charges in
mixed ionic-covalent compounds such as the HTSC's, because the charge
transfer from the cations to the anions is not complete as in the
entirely ionic case. Finally, applying the pair-potential approximation
we get for the total energy
\begin{equation}\label{2}
E(R,\zeta) = \sum_{\vc{a},\alpha} E_\alpha^\vc{a}(\zeta)
+\frac{1}{2}\sum_{(\vc{a},\alpha)\neq(\vc{b},\beta)}
\Phi_{\alpha\beta}
\left(\vc{R}^\vc{b}_\beta-\vc{R}^\vc{a}_\alpha,\zeta\right).
\end{equation}
The energy $E$ depends on both the configuration of the ions $\{R\}$
and the electronic (charge) degrees of freedom (EDF) $\{\zeta\}$ of the
charge density, i.e., $\{Q_{\lambda}\}$ and $\{\vc{p}_{\alpha}\}$ in
Eq. \eqref{1}. $E^\vc{a}_{\alpha}$ are the energies of the single ions.
$\vc{a}$, $\vc{b}$ denote the elementary cells and $\alpha,\,\beta$ the
corresponding sublattices. The second term in Eq. \eqref{2} is the
interaction energy of the system, expressed in terms of
\textit{anisotropic} pair interactions $\Phi_{\alpha\beta}$. Both
$E^\vc{a}_{\alpha}$ and $\Phi_{\alpha\beta}$ in general depend upon
$\zeta$ via $\rho_{\alpha}$ in Eq. \eqref{1}. The pair potentials in
Eq. \eqref{2} can be separated into long-ranged Coulomb contributions
and short-ranged terms as follows:
\begin{align}\nonumber
\Phi_{\alpha\beta}(\vc{R},\zeta) =& \frac{\mathcal{Z}_\alpha \mathcal{Z}_\beta}{R}
-(\mathcal{Z}_\alpha \vc{p}_\beta + \mathcal{Z}_\beta \vc{p}_\alpha)\cdot\frac{\vc{R}}{R^3}
+\frac{\vc{p}_\alpha\cdot\vc{p}_\beta}{R^3}\\\label{3}
&-3\frac{(\vc{p}_\alpha\cdot\vc{R})(\vc{R}\cdot\vc{p}_\beta)}{R^5}
+ \widetilde{\Phi}_{\alpha\beta}(\vc{R},\zeta),
\end{align}
\begin{align}\nonumber
\widetilde{\Phi}_{\alpha\beta}(\vc{R},\zeta) =& K_\alpha U_\beta(\vc{R},\zeta)
+ K_\beta U_\alpha(\vc{R},\zeta)\\\label{4}
&+ W_{\alpha\beta}(\vc{R},\zeta) + G_{\alpha\beta}(\vc{R},\zeta).
\end{align}
The first term in Eq. \eqref{3} describes the long-ranged ion-ion, the
second the dipole-ion and the third and fourth term the dipole-dipole
interaction. ${\cal{Z}}_{\alpha}$ and ${\cal{Z}}_{\beta}$ are the
variable charges of the ions in case the CF's are excited. The latter
reduce to the ionic charges for rigid ions. $K_{\alpha}$ and
$K_{\beta}$ are the charges of the ion cores. The remaining term in Eq.
\eqref{3} given in Eq. \eqref{4} represents the short-ranged
interactions. Detailed expressions for these interactions and the
procedure for the calculation can be found in Ref. \onlinecite{30}.

From the adiabatic condition
\begin{equation}\label{10}
\frac{\partial E(R,\zeta)}{\partial \zeta} = 0
\end{equation}
an expression for the atomic force constants, and accordingly the
dynamical matrix in harmonic approximation can be derived:
\begin{align}\label{11}\nonumber
t_{ij}^{\alpha\beta}(\vc{q}) &=
\left[t_{ij}^{\alpha\beta}(\vc{q})\right]_\text{RIM}\\ &-
\frac{1}{\sqrt{M_\alpha M_\beta}} \sum_{\kappa,\kappa'}
\left[B^{\kappa\alpha}_i(\vc{q}) \right]^{*} \left[C^{-1}(\vc{q})
\right]_{\kappa\kappa'} B^{\kappa'\beta}_j(\vc{q}).
\end{align}
The first term on the right hand side denotes the contribution from the
RIM. $M_{\alpha}$, $M_{\beta}$ are the masses of the ions and $\vc{q}$
is a wave vector from the first BZ.

The quantities $\vc{B}(\vc{q})$ and $C(\vc{q})$ in Eq. \eqref{11}
represent the Fourier transforms of the electronic coupling
coefficients as calculated from the energy in Eq. \eqref{2}, or the
pair potentials in Eqs. \eqref{3}, \eqref{4}, respectively:
\begin{align}\label{12}
\vc{B}_{\kappa\beta}^{\vc{a}\vc{b}} &= \frac{\partial^2
E(R,\zeta)}{\partial \zeta_\kappa^\vc{a} \partial R_\beta^\vc{b}},
\\\label{13} C_{\kappa\kappa'}^{\vc{a}\vc{b}} &= \frac{\partial^2
E(R,\zeta)}{\partial \zeta_\kappa^\vc{a} \partial
\zeta_{\kappa'}^\vc{b}}.
\end{align}
$\kappa$ denotes the EDF (CF and DF in the present model, see Eq.
\eqref{1}) in an elementary cell. The $\vc{B}$ coefficients describe
the coupling between the EDF and the displaced ions (bare
electron-phonon coupling), and the coefficients $C$ determine the
interaction between the EDF. The charge and dipole fluctuations
$Q_\lambda$ and $\vc{p}_\alpha$ do not appear explicitly in the
dynamical matrix. Only the derivatives of the energy with respect to
the EDF are needed in order to calculate the coefficients $\vc{B}$ and
$C$, respectively, and these derivatives have to be performed at
$Q_\lambda=0$, $\vc{p}_\alpha=\vc{0}$. The phonon frequencies
$\omega_{\sigma}(\vc{q})$ and the corresponding eigenvectors
$\vc{e}^{\alpha}(\vc{q}\sigma)$ of the modes $(\vc{q}\sigma)$ are
obtained from the secular equation for the dynamical matrix in Eq.
\eqref{11}, i.e.
\begin{equation}\label{14}
\sum_{\beta,j} t_{ij}^{\alpha\beta}(\vc{q})e_j^\beta(\vc{q}) =
\omega^2(\vc{q}) e_i^\alpha(\vc{q}).
\end{equation}
Equations \eqref{11}-\eqref{14} are generally valid and, in particular,
are independent of the specific model for the decomposition of the
perturbed density in Eq. \eqref{1} and the pair approximation Eq.
\eqref{2} for the energy. The lenghty details of the calculation of the
coupling coefficients $\vc{B}$ and $C$ cannot be reviewed in this
paper. They are presented in Ref. \onlinecite{30}. In this context we
remark that the coupling matrix $C_{\kappa\kappa'}(\vc{q})$ of the
EDF-EDF interaction, whose inverse appears in Eq. \eqref{11} for the
dynamical matrix, can be written in matrix notation as
\begin{equation}\label{15}
C = \Pi^{-1} + \widetilde{V}.
\end{equation}
$\Pi^{-1}$ contains the kinetic part to the interaction $C$ and
$\tilde{V}$ the Hartree and exchange-correlation contribution. $C^{-1}$
needed for the dynamical matrix and the EPI is closely related to the
(linear) density response function (matrix) and to the inverse
dielectric function (matrix) $\varepsilon^{-1}$, respectively. Only
very few attempts have been made to calculate the phonon dispersion and
the EPI of the HTSC's using the linear response method in form of
density functional perturbation theory (DFPT) within LDA, Refs.
\onlinecite{35,36,37}. These calculations correspond to calculating
$\Pi$ and $\tilde{V}$ in DFT-LDA and for the \textit{metallic} state
only. On the other hand, in our microscopic modeling DFT-LDA-SIC
calculations are performed for the various densities in Eq. \eqref{1}
in order to obtain the coupling coefficients $\vc{B}$ and $\tilde{V}$.
Including SIC is particularly important for localized orbitals such as
Cu$3d$ in the HTSC's. SIC as a correction for a single particle term is
not a correlation effect, which per definition cannot be described in a
single particle theory, but SIC is important for contracting in
particular the localized Cu$3d$ orbitals. Our theoretical results for
the phonon dispersion of $p$-doped materials, Refs.
\onlinecite{19,21,24}, which compare well with the experiments,
demonstrate that the approximative calculation of the coupling
coefficients in our approach is sufficient, even for the localized
Cu$3d$ states. Written in matrix notation we get for the density
response matrix the relation
\begin{equation}\label{16}
C^{-1} = \Pi(1+\widetilde{V}\Pi)^{-1} \equiv \Pi \varepsilon^{-1},
\hspace{.7cm} \varepsilon = 1 + \widetilde{V}\Pi.
\end{equation}
The CF-CF submatrix of the matrix $\Pi$ can approximatively be
calculated from a TBA of a single particle electronic bandstructure. In
this case the \textit{static} electronic polarizability $\Pi$ in
tight-binding representation reads
\begin{align}\nonumber
\Pi_{\kappa\kappa'}&(\vc{q},\omega=0) = -\frac{2}{N}\sum_{n,n',\vc{k}}
\frac{f_{n'}(\vc{k}+\vc{q})
-f_{n}(\vc{k})}{E_{n'}(\vc{k}+\vc{q})-E_{n}(\vc{k})} \times
\\\label{17} &\times \left[c_{\kappa n}^{*}(\vc{k})c_{\kappa
n'}(\vc{k}+\vc{q}) \right] \left[c_{\kappa' n}^{*}(\vc{k})c_{\kappa'
n'}(\vc{k}+\vc{q}) \right]^{*}.
\end{align}
$f$, $E$ and $c$ in Eq. \eqref{17} are the occupation numbers, the
single-particle energies and the expansion coefficients of the Bloch
functions in terms of tight-binding functions. The self-consistent
change of an EDF on an ion induced by a phonon mode $(\vc{q}\sigma)$
with frequency $\omega_{\sigma}(\vc{q})$ and eigenvector
$\vc{e}^{\alpha}(\vc{q}\sigma)$ can be derived in the form
\begin{align}\nonumber
\delta\zeta_\kappa^\vc{a}(\vc{q}\sigma) &= \left[-\sum_\alpha
\vc{X}^{\kappa\alpha}(\vc{q})\vc{u}_\alpha(\vc{q}\sigma)\right]
e^{i\vc{q}\vc{R}_\kappa^\vc{a}} \\
&\equiv \delta\zeta_\kappa(\vc{q}\sigma)e^{i\vc{q}\vc{R}^\vc{a}}, \label{18}
\end{align}
with the displacement of the ions
\begin{equation}\label{19}
\vc{u}_\alpha^{\vc{a}}(\vc{q}\sigma) =
\left(\frac{\hbar}{2M_\alpha\omega_\sigma(\vc{q})}
\right)^{1/2}\vc{e}^\alpha(\vc{q}\sigma)e^{i\vc{q}\vc{R}^\vc{a}}
\equiv \vc{u}_\alpha(\vc{q}\sigma)e^{i\vc{q}\vc{R}^\vc{a}}.
\end{equation}
The self-consistent response per unit displacement of the EDF in Eq.
\eqref{18} is calculated in linear response theory as
\begin{equation}\label{20}
\vc{X}(\vc{q}) = \Pi(\vc{q})\varepsilon^{-1}(\vc{q})\vc{B}(\vc{q}) =
C^{-1}(\vc{q})\vc{B}(\vc{q}).
\end{equation}
A measure of the strength of the EPI for a certain phonon mode
$(\vc{q}\sigma)$ is provided by the change of the self-consistent
potential in the crystal felt by an electron at some space point
$\vc{r}$ in this mode, i.e. $\delta
V_{\text{eff}}(\vc{r},\vc{q}\sigma)$. Averaging this quantity with the
corresponding density form factor $\rho_{\kappa}(\vc{r} -
\vc{R}^\vc{a}_{\kappa})$ at the EDF located at
$\vc{R}^\vc{a}_{\kappa}$, we obtain
\begin{equation}\label{21}
\delta V_\kappa^\vc{a}(\vc{q}\sigma) = \int dV
\rho_\kappa(\vc{r}-\vc{R}_\kappa^\vc{a}) \delta
V_\text{eff}(\vc{r},\vc{q}\sigma).
\end{equation}
This gives a measure for the strength of the EPI in the mode
$(\vc{q}\sigma)$ mediated by the EDF considered. For an expression of
$\delta V_{\kappa}^\vc{a}(\vc{q}\sigma)$ in terms of the coupling
coefficients given in Eqs. \eqref{12} and \eqref{13}, see Ref.
\onlinecite{19}. From our calculations for LaCuO large values for
$\delta V_{\kappa}^\vc{a}(\vc{q}\sigma)$ are found, in particular for
the OBSM phonon anomalies and even larger for the nonadiabatic $c$-axis
phonons in the metallic phase, mixing with a plasmon.

A nonadiabatic approach is necessary for a description of the
interlayer phonons and the charge response within a small region around
the $c$ axis, Refs. \onlinecite{19,38}.

Finally, it may be useful to outline, how the perturbed charge density
in Eq. \eqref{1} is related to the Fourier transform of the response
quantity $\vc{X}(\vc{q})$ in Eq. \eqref{20} and thus to $\Pi$ and
$\varepsilon$ describing the dielectric response of a specific
material.

The change of the electronic density $\rho(\vc{r})$, depending on the
ionic configuration $\{R\}$ and the EDF $\{\zeta\}$, induced by a unit
displacement of an ion at $\vc{R}^\vc{a}_\alpha$ is given by the
\textit{vector field}
\begin{equation}\label{31}
\vc{P}^\vc{a}_\alpha(\vc{r}) \equiv
\frac{d\rho(\vc{r})}{d\vc{R}^\vc{a}_\alpha} =
\frac{\partial\rho(\vc{r})}{\partial\vc{R}^\vc{a}_\alpha} +
\sum_{\vc{b}\,\kappa} \frac{\partial \rho(\vc{r})}{\partial
\zeta^\vc{b}_\kappa} \frac{\partial \zeta^\vc{b}_\kappa}{\partial
\vc{R}^\vc{a}_\alpha}.
\end{equation}
By differentiating the adiabatic condition in Eq. \eqref{10} with
respect to $\vc{R}^\vc{a}_\alpha$ and using the Hohenberg-Kohn
functional of DFT for the energy the second term in Eq. \eqref{31} can
be put into the form\cite{20}
\begin{equation}\label{32}
\vc{P}^\vc{a}_{\alpha,\text{d}}(\vc{r}) \equiv -
\sum_{\vc{b}\,\kappa}\rho_\kappa(\vc{r}-\vc{R}^\vc{b}_\kappa)
\vc{X}_{\kappa\,\alpha}^{\vc{b}\,\vc{a}},
\end{equation}
where the form factors of the EDF's $\kappa$ are defined by
\begin{equation}\label{33}
\rho_\kappa(\vc{r}-\vc{R}^\vc{b}_\kappa) =
\frac{\partial\rho(\vc{r})}{\partial\zeta_\kappa^\vc{b}}
\end{equation}
and $\vc{X}_{\kappa\,\alpha}^{\vc{0}\,\vc{a}}$ is the Fourier transform
of $\vc{X}^{\kappa\alpha}(\vc{q})$ in Eq. \eqref{18}. Equation
\eqref{31} gives a decomposition of the charge response into an
explicit (rigid) change of the density, defined by
\begin{equation}\label{34}
\vc{P}^\vc{a}_{\alpha,\text{r}}(\vc{r}) \equiv
\frac{\partial\rho(\vc{r})}{\partial\vc{R}^\vc{a}_\alpha},
\end{equation}
and the distortion contribution
$\vc{P}^\vc{a}_{\alpha,\text{d}}(\vc{r})$ from Eq. \eqref{32}.
Physically, the explicit contribution
$\vc{P}^\vc{a}_{\alpha,\text{r}}(\vc{r})$ describes the part of the
density which rigidly follows the motion of the ions and represents the
local EPI effects. On the other hand,
$\vc{P}^\vc{a}_{\alpha,\text{d}}(\vc{r})$ describes the way, how the
charge density is distorted in response to the displacement of an ion.
It represents the nonlocal EPI effects.

Such an \textit{unique} decomposition of the perturbed density into a
rigid and nonrigid (distortion) contribution has been shown to be
rigorously valid in linear response theory, see the discussion of the
\textit{Quasi-Ion concept} in Ref. \onlinecite{31}. In case of a
diagonal response function (matrix) in reciprocal space the distortion
contribution to the charge response vanishes and only the rigid
Quasi-Ion (Pseudoatom) contribution related to local EPI
survives\cite{31}. Especially, this is true for homogeneous electrongas
screening. On the other hand, in a strongly inhomogeneous electronic
system like the cuprates the density matrix in reciprocal space, is of
course, nondiagonal and apparently there are important distortion
contributions associated with the strong nonlocal EPI effects which add
to the charge response. In our modeling these distortion effects are
described in terms of ionic CF's and DF's, respectively. This is an
adequate approximation because of the specific solid-state chemistry of
these compounds and has been confirmed so far by our calculations for
$p$-doped materials and also for $n$-doped cuprates in this work.

In the present work the rigid part is approximated by the rigid-ion
density $\rho^0_\alpha$ in Eq. \eqref{1}, which works well for
materials with a strong component of ionic binding and the distortion
part by the two remaining terms with the corresponding form factors of
the CF's and DF's as given in this equation. The latter yield the shape
of the change in the density related to the CF's and DF's,
respectively.

A central problem of superconductivity in the cuprates is as to how the
metallic state evolves from the Mott-insulator upon doping. In
particular, there is no consense for a realistic description of the
underdoped region. Thus, there remains the important question as to how
to discriminate between the charge response of the metallic and
insulating state of the HTSC's, respectively. Specifically, the latter
cannot be obtained for example within the LDA and a realistic
quantitative description to calculate the \textit{irreducible} (proper)
polarization part $\Pi_{\kappa\kappa'}(\vc{q})$ for the HTSC's is not
available. However, a general criterion which \textit{necessarily}
requires a \textit{multi-orbital} approach  follows from the different
analytic behaviour of the orbital resolved polarizability in the
longwavelength limit $(\vc{q} \to \vc{0})$ in both phases \cite{19,
20}.

In the metallic phase the electronic partial density of states (PDOS)
at the Fermi level $Z_{\kappa}(\varepsilon_{F})$ is related to the
polarizability matrix for $(\vc{q} \to \vc{0})$ according to
\begin{equation}\label{22}
\sum_{\kappa'} \Pi_{\kappa\kappa'}(\vc{q}\rightarrow\vc{0}) =
Z_\kappa(\varepsilon_\text{F}),
\end{equation}
and the total density of states (DOS) at energy $\varepsilon$ is given
by
\begin{equation}\label{23}
Z(\varepsilon) = \sum_{\kappa} Z_\kappa(\varepsilon).
\end{equation}
On the other hand, for the insulating state we obtain the sum rules
\begin{equation}\label{24}
\sum_{\kappa'} \Pi_{\kappa\kappa'}(\vc{q}\rightarrow\vc{0}) =
\mathcal{O}(q)
\end{equation}
and
\begin{equation}\label{25}
\sum_{\kappa\kappa'} \Pi_{\kappa\kappa'}(\vc{q}\rightarrow\vc{0}) =
\mathcal{O}(q^2).
\end{equation}
The sum $\sum_{\kappa\kappa'}\Pi_{\kappa\kappa'}(\vc{q} \to \vc{0})$ is
equal to $\rho^{2}K$ with $\rho$ the average density and $K$ the
compressibility of the electronic system. The latter provides a measure
of the gap in the electronic spectrum because $K$ vanishes as a
function of the chemical potential in the gap region.

Equations \eqref{22}-\eqref{25}, respectively, can be considered as an
orbital resolved closed form to describe the metal-insulator transition
in terms of the polarizability or the compressibility of the electronic
system, respectively. This is a primary tool to characterize the
corresponding ground state. Such orbital selective sum rules are
particularly useful in the proximity to a Mott insulating phase where
different from the band insulator the internal degrees of freedom,
orbital (and spin), approximatively survive. In order to fulfil the sum
rules above, in case of the insulating state, in contrast to the
metallic state, off-diagonal elements of the polarizability matrix
describing nonlocal polarization processes necessarily must occur and
interfere in order to correlate the charge response in such a way that
Eqs. \eqref{24} and \eqref{25} are satisfied.

Physically this is related to the fact that in an insulator a
perturbation, e.g., a change of the electron-ion potential, is only
incompletely screened. Consequently, the selfconsistent change of the
potential at the orbital degree of freedom $\kappa'$ nonlocally
contributes to the CF's at the orbital $\kappa \not= \kappa'$ in the
unit cell. This is quite different from the metallic case where the
diagonal elements of $\Pi_{\kappa\kappa'}$ dominate and off-diagonal
elements can be neglected in most cases, i.e., only the PDOS
$Z_{\kappa}(\varepsilon_{F})$ contribute.

Approaching the delocalization-localization transition from the
metallic region when $p$-doping is decreased according to our modeling
of the cuprates the Cu$3d$ component of the wave function is admitted
to become incompressible, insulator-like in the underdoped state, i.e.,
in terms of the sum rule we have \cite{23}
\begin{equation}
\sum\limits_{\kappa^{'}} \Pi_{\kappa\kappa^{'}}(\vc{q} \to \vc{0}) = \left\{ \begin{array}{lcl}
{\cal{O}}(q) & , & \text{Cu$3d$}\\
\mbox{}\\
\widetilde{Z}_{\kappa}(\varepsilon_{F}) & , & \text{O$2p$} \end{array}
\right. \label{26}
\end{equation}
for the $p$-doped materials. In general, an orbital of type $\kappa$ is
defined to be compressible or incompressible, respectively, in case the
sum rule in Eq. \eqref{22} for a metal or Eq. \eqref{24} for an
insulator is satisfied for the orbital in question.

In contrast to the $p$-doped case we propose in this work for the
$n$-doped materials in the underdoped state
\begin{equation}
\sum\limits_{\kappa^{'}} \Pi_{\kappa\kappa^{'}}(\vc{q} \to \vc{0}) = \left\{ \begin{array}{lcl}
{\cal{O}}(q) & , & \text{O$2p$}\\
\mbox{}\\
\widetilde{Z}_{\kappa}(\varepsilon_{F}) & , & \text{otherwise}
\end{array} \right. , \label{27}
\end{equation}
i.e., the O$2p$ component becomes incompressible, localized while the
Cu$3d$ and Cu$4s$ component remains metallic, delocalized as in the
optimally doped metallic phase but with a \textit{renormalized} density
of states $\widetilde{Z}_{\kappa}(\varepsilon_{F})$. We call a state
defined by the sum rules from Eqs. \eqref{26} and \eqref{27},
respectively, a ''\textit{strange metal}'', because not all orbitals in
the system are compressible as in the normal metallic state related to
the sum rule from Eqs. \eqref{22}, \eqref{23} representing in general a
Fermi liquid.

Nevertheless, the total compressibility is never zero in such a strange
metallic state and a real space organization for the low lying charge
excitations is achieved in a complementary way in both cases. Quite
generally, the incompressible regions compete with overall metallic
behaviour and with superconductivity. In our modeling of the underdoped
phase of the HTSC's the PDOS for Cu$3d$ is suppressed at the Fermi
level in case of $p$-type HTSC's and for O$2p$ in the case of $n$-type
materials. So, we have for $p$-doped materials an orbital selective
compressible, metallic charge response for the holes in the O$2p$
orbitals with a renormalized PDOS. For $n$-doped materials the charge
response is metallic at the Cu sites but nonmetallic for the O$2p$
orbitals simulating the absence of a hole contribution in the
underdoped non-superconducting material. Altogether, we have a sudden
loss in the density of states (\textit{pseudogap}) at the Fermi level
in both cases. Thus, the quasiparticle picture
($Z_{\kappa}(\varepsilon_{F}) \not= 0$ for \textit{all} $\kappa$)
consistent with our modeling of the metallic-, superconducting state
from optimal to overdoping no longer holds for the underdoped regime.
In the latter, we have a renormalized density of states at
$\varepsilon_{F}$ which most likely is due to an orbital-dependent
distortion of the bandstructure and a corresponding reconstruction of
the large FS to some smaller one. Finally, in the optimally $n$-doped
metallic state the formerly incompressible O$2p$ orbital is allowed to
become compressible metallic too as in $p$-doped LaCuO and holes can
now contribute to the metallic properties and superconductivity in
$n$-typed materials consistent with the development of an LDA-like FS
of hole type in NdCuO. In summary, in our approach the different ground
states of the cuprates are distinguished by certain orbital resolved
incompressibility- compressibility transitions related to corresponding
transformations of the FS.

Off-site CF's at the center of the planar oxygen squares can be
introduced additionally, simulating specific covalence effects as
described in Ref. \onlinecite{24} in order to improve selectively the
phonon branch with the scissor mode at the $X$ point. In this mode the
in-plane oxygens vibrate perpendicular to the CuO bonds. These specific
off-site CF's virtually have no effect on the other phonon branches.

By our modeling of the underdoped state in both cases a \textit{partial
ordering} of the conducting carriers in the CuO plane is obtained. With
regard to our modeling of the metallic state for optimally to overdoped
HTSC's  such an ergodicity breaking delocalization-localization
transition in the underdoped state in terms of the compressibility
means a compartmentalization of configuration space. This means that
some parts of direct space cannot be approached or are hardly
accessible to the charge carriers in the \textit{pseudogap state}. On
the other hand, this should be accompanied by a reciprocal
compartmentalization of momentum space, according to the uncertainity
relationship $\Delta x\,\Delta k \sim 1$. Indeed, the ARPES studies in
the underdoped cuprates, Refs. \onlinecite{25,26,27,28}, point to such
a compartmentalization of reciprocal space. The corresponding
transformation of a large FS into Fermi surface arcs or small Fermi
surface pockets, respectively, would result in a loss of density of
states at the Fermi level (pseudogap) consistent with our modeling.

\section{PHONON DISPERSION OF NdCuO IN THE INSULATING AND METALLIC PHASE}\label{sec_dispersions}
\subsection{Phonon dispersion in the insulating state}
\subsubsection{The ionic reference system - RIM}
\begin{figure}
\includegraphics[width=\linewidth]{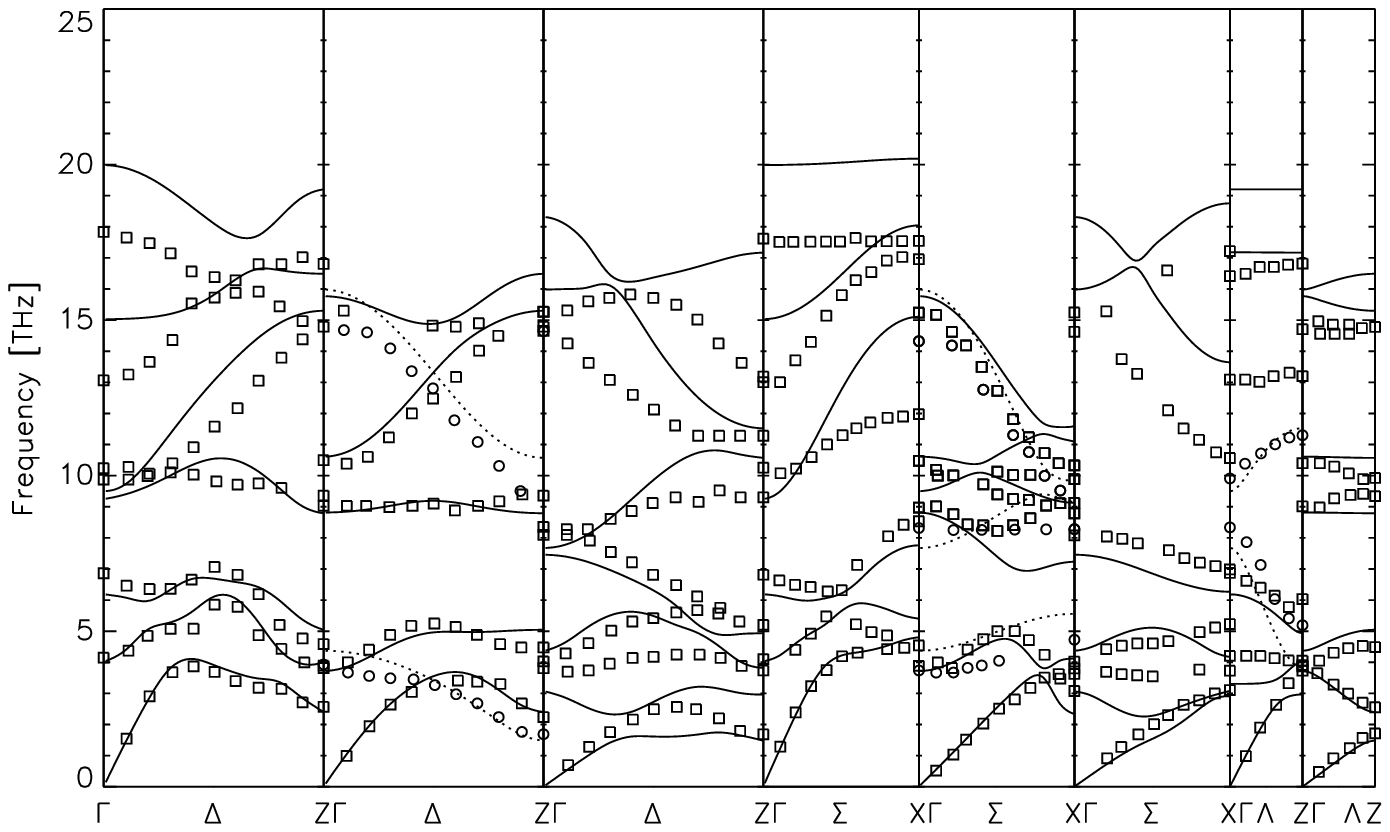}
\caption{Calculated phonon dispersion of NdCuO in the main symmetry
directions $\Delta \sim (1,0,0)$, $\Sigma \sim (1,1,0)$, and $\Lambda
\sim (0,0,1)$ as obtained for the rigid-ion model explained in the
text. The experimental data for insulating (undoped) NdCuO are taken
from Ref. \onlinecite{39}. The diverse symbols representing the
experimental results from inelastic neutron scattering (INS) mark the
different irreducible representations (ID). The arrangement of the
panels from left to right is as follows:
$\Delta_{1}|\Delta_{2}\,(\cdots,\,\circ)$,
$\Delta_{4}\,(-\!\!\!-,\,\Box)|\Delta_{3}|\Sigma_{1}|\Sigma_{2}\,(\cdots
,\,\circ)$,
$\Sigma_{4}\,(-\!\!\!-,\,\Box)|\Sigma_{3}|\Lambda_{1}\,(-\!\!\!-,\,\Box)$,
$\Lambda_{2}\,(\cdots,\,\circ)|\Lambda_{3}|$. The frequencies are in
units of THz.}\label{FIG01}
\end{figure}

For a definitive discussion of the phonon renormalization induced by
the nonlocal EPI effects of CF and DF type, mediated by the second term
in the dynamical matrix from Eq. \eqref{11}, a quantitative reference
model for the calculation of the phonon dispersion based essentially on
the ionic component of binding is necessary. Such a model sketched in
Sec. \ref{sec_theory}, representing approximately the local EPI
effects, is provided by the \textit{ab initio} rigid-ion model,
extended via covalent ion softening and scaling of the short-ranged
part of certain pair potentials. The results for the phonon dispersion
of NdCuO along the main symmetry directions $\Delta \sim (1,0,0)$,
$\Sigma \sim (1,1,0)$ and $\Lambda \sim (0,0,1)$ are shown in Fig.
\ref{FIG01} where it is compared with the experimental results from
inelastic neutron scattering (INS) for the insulating state \cite{39}.
The static effective charges found for the model are Nd$^{2.35+}$,
Cu$^{1.22+}$, O$_{xy}^{1.42-}$, O$_{z}^{1.54-}$, which should be
compared with the nominal ionic charges Nd$^{3+}$, Cu$^{2+}$,
O$_{xy}^{2-}$, O$^{2-}_{z}$. The difference between these two sets of
charges indicates the degree of ion softening. We see from Fig.
\ref{FIG01} that the calculated curves partially represent the
experimental results, already quite well. However, the high-frequency
optical phonon modes (e.g. the E$_{u}$-modes at the $\Gamma$ point) are
considerably overestimated. This points to a missing electronic
polarization mechanism in the RIM. The strong ionic component of
binding indicates that nonlocal DF's should be an important electronic
degree of freedom in the charge response.

We note that the RIM leading to the dispersion in Fig. \ref{FIG01} also
gives reasonable structural data. So we obtain for the energy-minimized
structure for the planar lattice constant $a = 4.155\AA$, for the
lattice constant along the $c$-axis $c = 12.747\AA$ and for the
internal position of the Nd ion $z(\text{Nd}) = 0.153$ in units of $c$.
The experimental values are $a = 3.945\AA$, $c = 12.171\AA$ and
$z(\text{Nd}) = 0.149$, Ref. \onlinecite{40}.

From our numerical calculations for a purely ionic RIM with nominal
ionic charges, not shown here, we find the following effects of
covalent softening and scaling on the phonon dispersion. Scaling of the
short-ranged part of the pair potentials has the qualitative effect to
stabilize all formerly unstable modes found in a purely ionic RIM.
Moreover, we find a decrease of the width of the spectrum towards the
width of the experimental result. Ion softening also decreases the
spectral width but hardly stabilizes the unstable modes. Altogether, a
considerable improvement of the calculated dispersion is achieved as
compared to the results from a purely ionic model  using nominal ionic
changes and neglecting scaling.

\subsubsection{RIM plus dipole fluctuations}
\begin{figure}
\includegraphics[width=\linewidth]{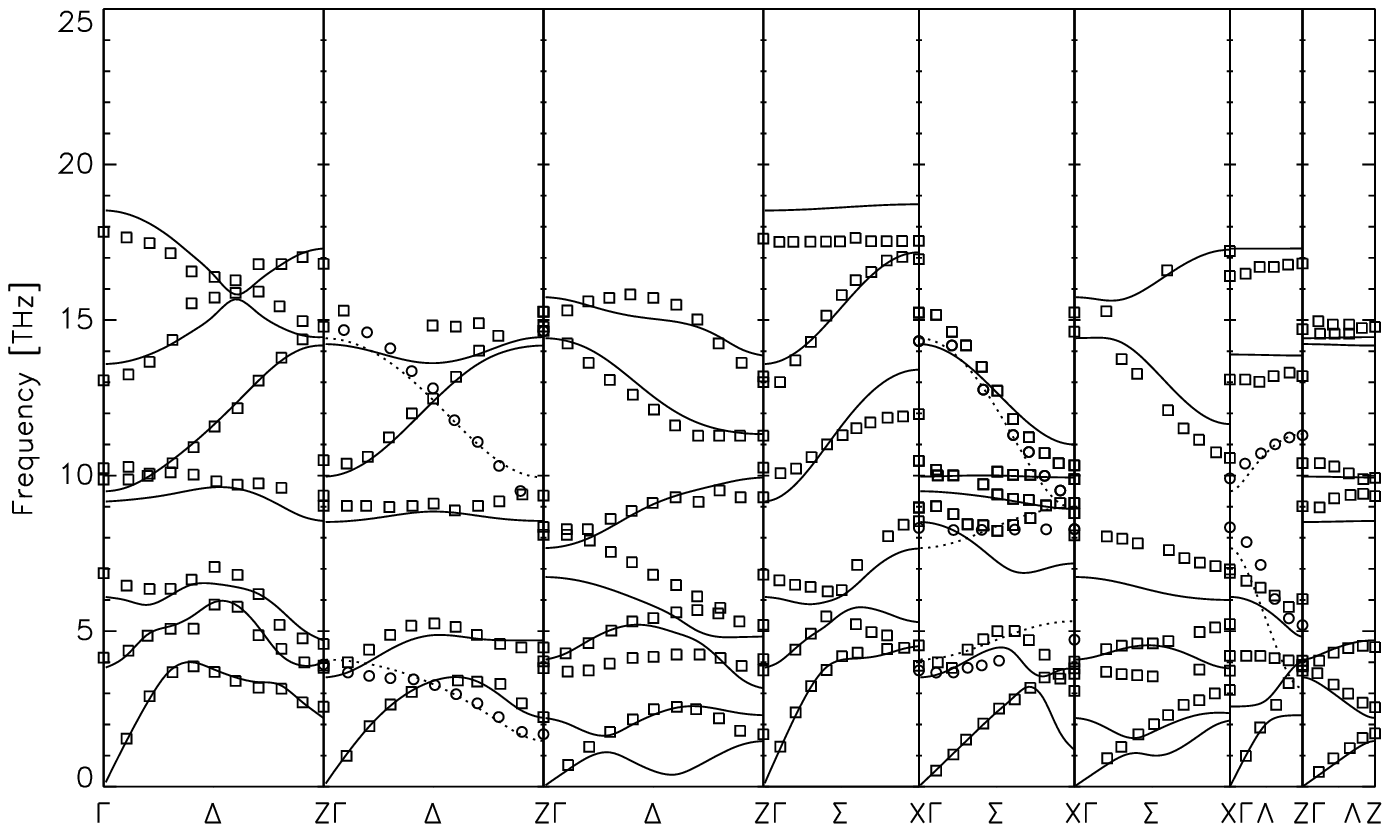}
\caption{Same as in Fig. \ref{FIG01} with the calculated results
obtained taking additionally anisotropic dipole fluctuations (DF's)
into account.}\label{FIG02}
\end{figure}

Likewise as in our calculations for the $p$-doped materials LaCuO
\cite{24}, YBaCuO \cite{41}, Bi-2201 and Bi-2212 \cite{42}, a
significant improvement of the calculated phonon dispersion as compared
to the RIM can be obtained also for NdCuO by introducing additionally
DF's into the modeling, see Fig. \ref{FIG02}. As in the calculations
for the $p$-doped materials the dipole polarizability turns out to be
\textit{anisotropic} for the different ions, i.e., the calculated ab
initio values of the dipole polarizability $\alpha$ as obtained from
the Sternheimer method for the isolated ions have in general to be
reduced in a suitable way in the crystal. The actual values used in the
calculations are listed in Table \ref{tabdipol} .This leads to a better
agreement of the phonon dispersion with the experiments. A comparison
of the calculated data in Fig. \ref{FIG02} with the experimental ones
shows an overall good agreement. This means that already the allowance
of DF's in the electronic screening process leads to reasonable results
for the phonon dispersion of insulating NdCuO reflecting its ionic
nature.

\begin{table}
\begin{tabular}{c|cccc}
    &  Cu &  O$_{xy}$  &  O$_z$  &  Nd   \\\hline\hline
ab-initio & 8.9 & 7.2 & 8.2 & 12.5 \\\hline
xy  & $30\%$ &  $30\%$   & 40$\%$  & 40$\%$\\
z   &40$\%$&  100$\%$  & 40$\%$ & 40$\%$\\
\end{tabular}
\caption{The calculated result for the dipole polarizability according
to the Sternheimer method\cite{30} in units of $a_\text{B}^3$ is given
in the first row for Cu$^{1.22+}$, O$_{xy}^{1.42-}$, O$_z^{1.54-}$ and
Nd$^{2.35+}$ . The anisotropic reduction of the polarizability for the
insulator is presented in the last two lines in percent. 100 percent
stands for the ab initio result.}\label{tabdipol}
\end{table}

Contrasting the calculated results of Fig. \ref{FIG01} with Fig.
\ref{FIG02} the important effect of the DF's on the disperion can be
figured out, in particular for the high-frequency optical modes. For
example, the frequencies of the longitudinal E$_{u}$ modes as well as
the corresponding LO-TO splittings are properly decreased.  The latter
come out too large in the RIM, but now are in agreement with the
experiments. As a minor deficit the width of the spectrum is still a
bit too large. The two $\Delta_{1}$-branches with the highest
frequencies overestimate the frequencies at the $\Gamma$ point (E$_{u}$
modes) by about 0.5 THz and the highest $\Sigma_{1}$ branch is about 1
THz too high in frequency.

\subsubsection{RIM plus dipole- and charge fluctuations}
In a further step besides DF's as in Fig. \ref{FIG02} CF's are
additionally introduced into the modeling according to the sum rule
from Eq. (\ref{24}) for the charge response of an insulator. As in the
case of LaCuO, Ref. \onlinecite{24}, CF's  are allowed at the Cu$3d$
and O$2p$ orbitals in the CuO plane.

\begin{figure}
\includegraphics[width=\linewidth]{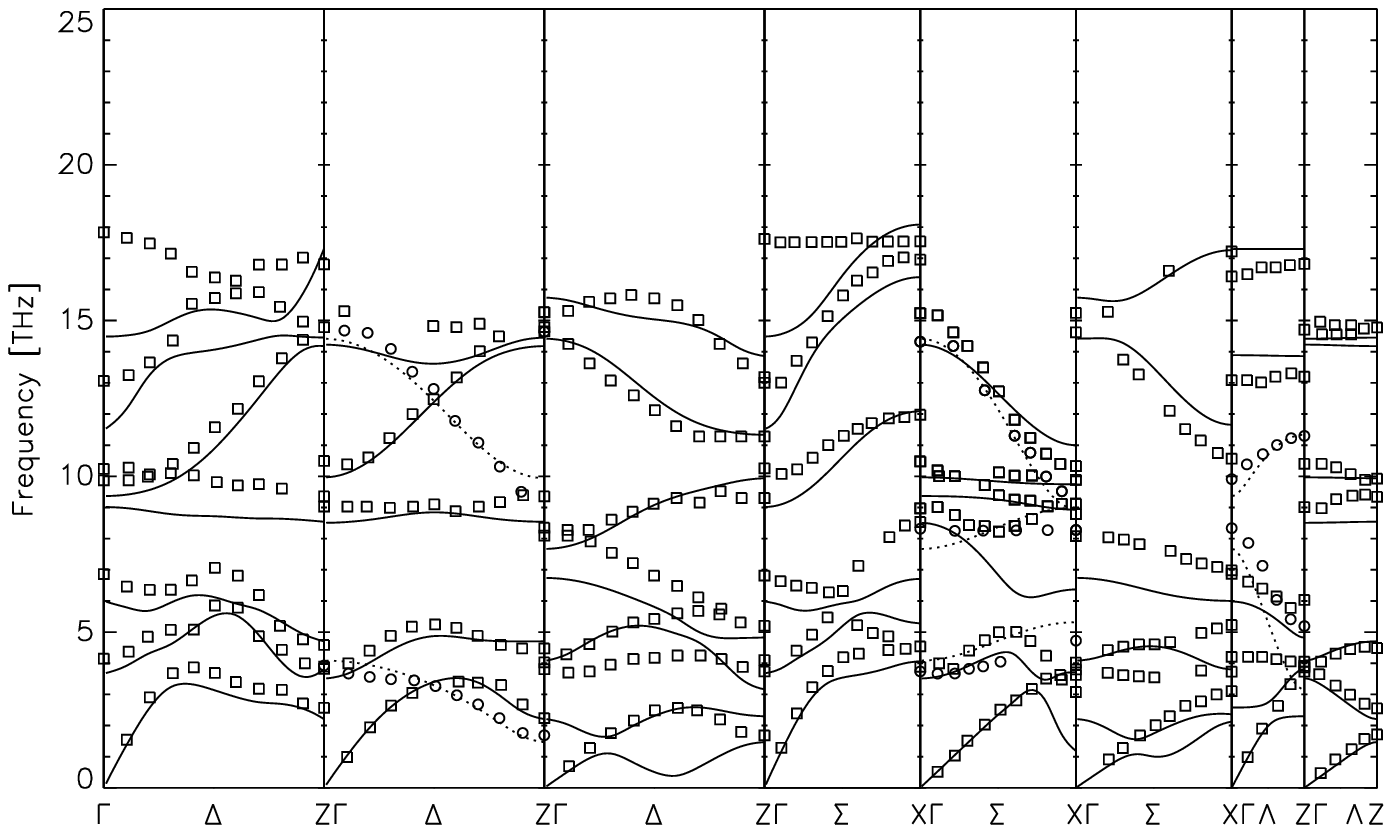}
\caption{Same as in Fig. \ref{FIG01} with the calculated results
achieved taking additionally DF's as in Fig. \ref{FIG02} and charge
fluctuations (CF's) according to earlier calculations for LaCuO, Ref.
\onlinecite{24}, into account.}\label{FIG03}
\end{figure}

In a first attempt we adopt the matrix elements of the electronic
polarizability $\Pi_{\kappa\kappa'}$ for the CF's from our calculations
for LaCuO \cite{24}, leading in the latter case to a good
representation of the INS data. The corresponding results for NdCuO of
such a modeling are displayed in Fig. \ref{FIG03}. As compared with
Fig. \ref{FIG02}, we recognize in contrast to the experiment a very
strong softening of the dispersion of the two characteristic
$\Delta_{1}$ and $\Sigma_{1}$ branches which display the anomalous
behaviour widely seen in the metallic phase of the HTSC's, see below.
In particular, at the $\Gamma$-point a very large decrease in frequency
is obtained for the high-frequency E$_{u}$-modes leading to a strong
disagreement with the experiment. From our numerical studies we find
that the reason for this disagreement is the large matrix elements for
the electronic polarizability $\Pi(\text{O}_{xy}2p)$ and
$\Pi(\text{O}_{x}2p-\text{O}_{y}2p)$ of the O$2p$ orbital adopted from
the case of LaCuO.

\begin{figure}
\includegraphics[width=\linewidth]{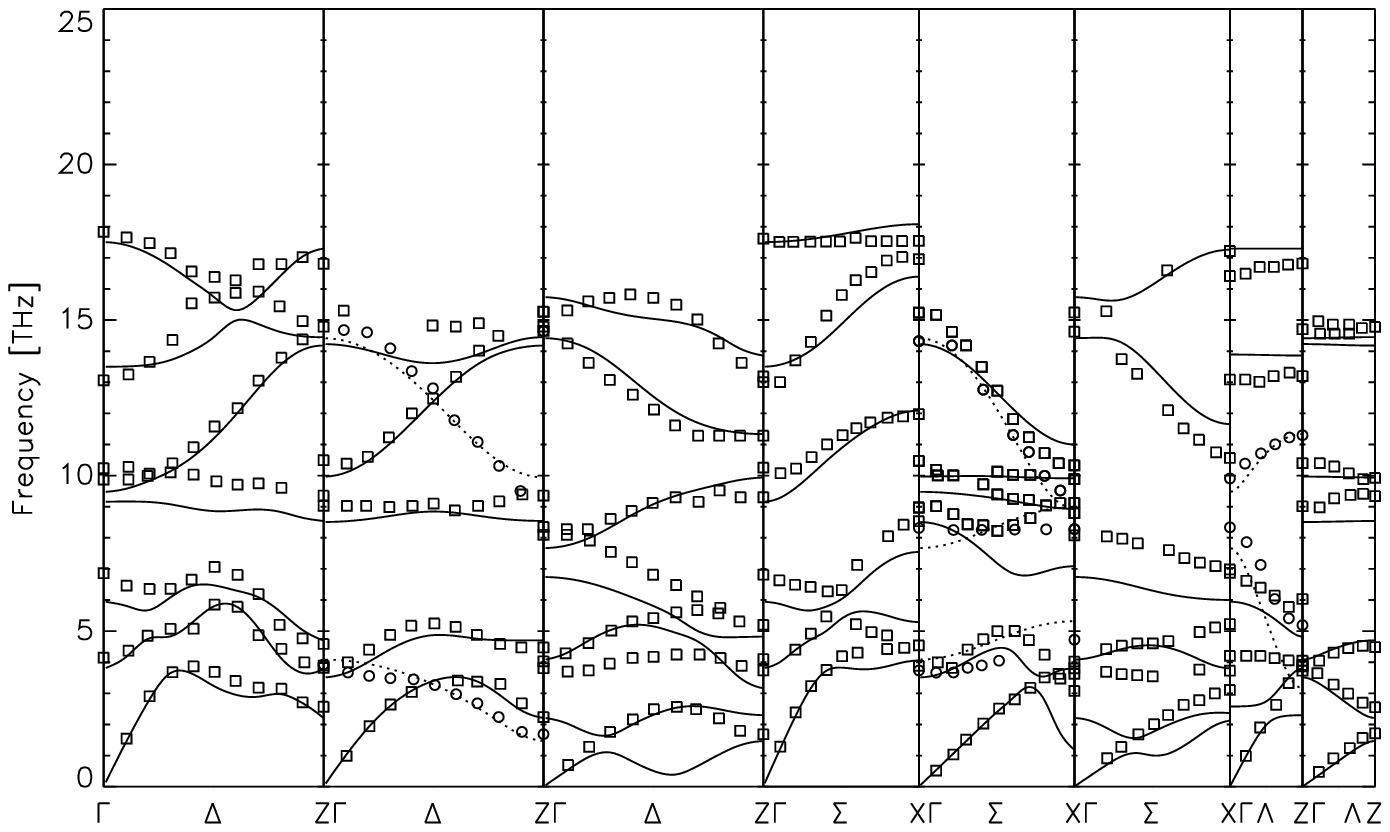}
\caption{Same as in Fig. \ref{FIG01} with the calculated results for
insulating NdCuO received taking additionally DF's as in Fig.
\ref{FIG02} and optimized CF's with a strongly reduced
CF-polarizability of the O$2p$ orbital for NdCuO into
account.}\label{FIG04}
\end{figure}

\begin{figure}%
\begin{minipage}{0.4\linewidth}%
\includegraphics[width=\linewidth]{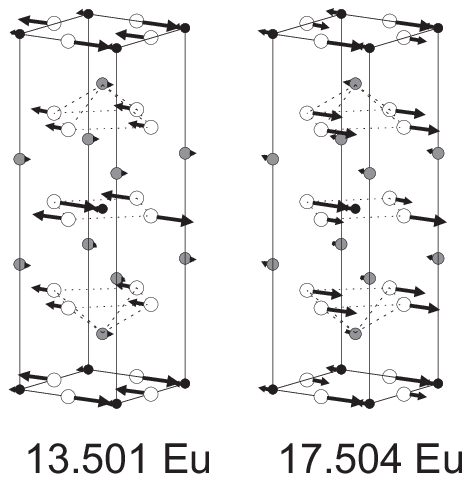}%
\end{minipage}%
\hspace{0.1\linewidth}%
\begin{minipage}{0.5\linewidth}%
 \caption{Displacement patterns of the two E$_{u}$ modes of CuO bond-stretching type
at the $\Gamma$ point with the highest frequencies for the model from
Fig. \ref{FIG04}. The frequency of the modes given below the patterns is in units of THz.}\label{FIG05}%
\end{minipage}%
\end{figure}

In order to obtain the result for the dispersion of insulating NdCuO
including DF's and CF's displayed in Fig. \ref{FIG04}, which is in good
agreement with the experiments, the polarizability of the O$2p$ orbital
concerning CF's has to be reduced to practically zero in NdCuO. Any
enhancement of the O$2p$-polarizability leads to a characteristic
softening of the two $\Delta_{1}$ and $\Sigma_{1}$ branches with
highest frequency towards the $\Gamma$-point in disagreement with
experiment. The corresponding high-frequency E$_{u}$ modes at the
$\Gamma$-point are of CuO bond-stretching type, see Fig. \ref{FIG05},
and are consequently very sensible to the CF-polarizability of the
O$2p$ orbital. Thus, we find from our phonon calculations a significant
different behaviour of the O$2p$ orbital in the charge response in
terms of CF's when comparing LaCuO with NdCuO. The nearly vanishing
CF-polarizability of O$2p$ in NdCuO points to stronger localization of
that orbital in NdCuO than in case of LaCuO and is an expression of an
enhancement of the ionic component of binding.

\subsection{Phonon dispersion in the metallic state}\label{sec_NCO_metall}
In order to describe the kinetic part of the charge response for the
metallic optimally doped (superconducting) phase of NdCuO we apply for
the modeling of $\Pi_{\kappa\kappa}$ the sum rule for the metallic
state  from Eq. \eqref{22}. Likewise as in our calculation for
optimally $p$-doped LaCuO \cite{24}, CF's are only permitted in the CuO
plane. In detail there are CF's of the Cu$3d$ and O$2p$ orbitals
admitted as in the insulating state. Moreover, CF's of the delocalized
Cu$4s$ orbital are additionally allowed. The latter, introduced via the
$\Pi$(Cu$4s$) matrix element of the polarizability, have been found in
the $p$-doped materials to be important for a correct description of
the characteristic phonon anomalies in the highest $\Delta_{1}$ and
$\Sigma_{1}$ branch, Refs. \onlinecite{19,21,23,24,43}. The growing
importance of the Cu$4s$ component in the wavefunction, which leads to
enhanced delocalized CF's at the Cu sites in the screening process, has
been related to a corresponding softening of the anomalous OBSM as
compared to calculations where the Cu$4s$ degree of freedom is
neglected, see e.g. Ref. \onlinecite{43}. So, from these calculations
additionally to Cu$3d$ and O$2p$ an increasing contribution of the
delocalized Cu$4s$ component to the metallic state of the HTSC's is
important to understand the charge response and the corresponding
phonon softening of the OBSM, at least at higher doping levels.

The importance of the Cu$4s$ orbital for a realistic description  of
the electronic structure of the HTSC's also has been pointed out in
Refs. \onlinecite{44,45}. In this work, the hopping range has been
identified as an essential material dependent parameter and the
intralayer hopping beyond nearest neighbours as well as interlayer
hopping proceeds via the Cu$4s$ orbital. It is further concluded that
materials with higher $T_{c,\text{max}}$ have larger hopping ranges and
in materials with highest $T_{c,\text{max}}$ the axial orbital,
essentially a hybrid between Cu$4s$, Cu$3d_{z^{2}-1}$ and apical oxygen
$2p_{z}$ ist almost pure Cu$4s$. Moreover, the importance of Cu$4s$ for
an accurate description of partial charge distributions in the HTSC's
is pointed out in Refs. \onlinecite{46,47}.

\begin{figure}
\includegraphics[width=\linewidth]{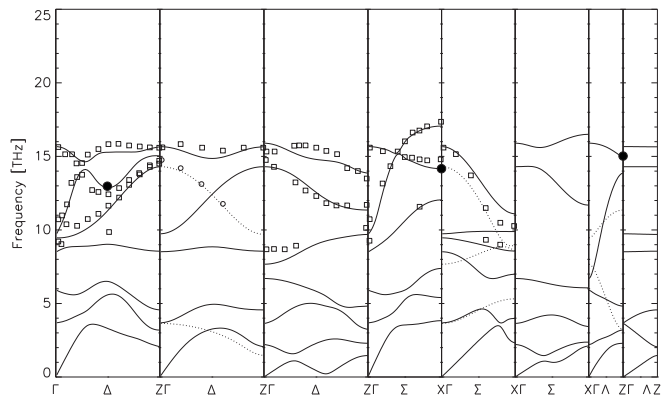}
\caption{Same as in Fig. \ref{FIG01} with the calculated results for
metallic NdCuO taking additionally DF's and CF's as discussed in the
text into account. The experimental INS results for optimally doped
NdCuO shown by the diverse symbols, are taken from Ref.
\onlinecite{18}. The three black dots ($\bullet$) in the figure
indicate the anomalous high-frequency oxygen bond-stretching modes
(OBSM) $\Delta_{1}/2$, $O_{B}^{X}$ and $O_{z}^{Z}$,
respectively.}\label{FIG06}
\end{figure}

In Fig. \ref{FIG06} we present our prediction of the phonon dispersion
for metallic NdCuO. We obtain a good agreement with the experimental
results as measured so far by INS, Ref. \onlinecite{18}. In particular,
the phonon anomalies are well represented. Our calculations support the
INS results which on her part to some extent deviate from the
corresponding inelastic x-ray scattering (INX) results, Refs.
\onlinecite{16,17}, in particular as far as the phonon anomalies are
concerned. The INS experiments find the anomalous oxygen half-breathing
mode (\dhalbe anomaly) at $\frac{2\pi}{a}(0.5,0,0)$ with a frequency of
12.5 THz, which is close to our calculated value of 12.868 THz, while
in the INX results this mode is preliminary identified at
$\frac{2\pi}{a}(0.2,0,0)$ with a frequency of about 12 THz. Our
calculated data in Fig. \ref{FIG06} for the two $\Sigma_{1}$ branches
with the highest frequencies are also in very good agreement with the
INS data while the INX results fall about 1.5 THz below the neutron
data for the lower of the two $\Sigma_{1}$ branches with the anomalous
planar oxygen breathing mode (\obx) at the $X$ point of the BZ.

The calculated results shown in Fig. \ref{FIG06} for the metallic phase
of NdCuO have been obtained with a model where the matrix elements
$\Pi_{\kappa\kappa}$ have been increased for the Cu$4s$ orbital but
strongly decreased for O$2p$ compared to our results for metallic
LaCuO, Ref. \onlinecite{21}. Thus, the tendency to a stronger
localization correlated with a much smaller CF-polarizability of the
O$2p$ orbitals in NdCuO, i.e. a $n$-doped material, as compared to
LaCuO, i.e. a $p$-doped material, as found already in the insulating
state ist also present in the metallic state.

An enhanced localization of O$2p$ in NdCuO tends to stabilize the
antiferromagnetic spin correlations in comparison to a $p$-doped
material like LaCuO, because the localization of the O$2p$ orbital is
favourable for superexchange. Such a trend is consistent with the fact
that the phase diagram of the cuprates is asymmetric with respect to
electron and hole doping. For the $n$-type materials the
antiferromagnetic phase extends much further with doping and competes
with superconductivity. On the other hand, allowing the O$2p$ orbitals
to become compressible, metallic with a significantly larger
CF-polarizability, i.e. more delocalized as in our modeling of the
$p$-doped HTSC's, makes the whole system more metallic, reduces
antiferromagnetic fluctuations and triggers superconductivity already
at a low doping level.

The enhanced Cu$4s$ CF-polarizability and the decreased O$2p$
CF-polarizability in a $n$-doped material like NdCuO as compared to a
$p$-doped one seems to be consistent with an \textit{electron-hole
asymmetry} introduced by the doping process. While in the $n$-doped
case materials electrons go to the hybridized orbitals of Cu$3d$ and
Cu$4s$ character during $p$-doping the holes are generated in the O$2p$
orbitals and enhance the CF-polarizability of the latter as suggested
by our calculations of the phonon dispersion.

Altogether, our calculations of the phonon dynamics in $p$-doped and
$n$-doped HTSC's point to a very different magnitude of the
CF-polarizability of the oxygen ions in these materials. Thus, one
might speculate that the CF-polarizability of oxygen should be
indirectly relevant for both, antiferromagnetic behaviour and
superconductivity. The related stronger localization of the O$2p$
orbital seems to favour in the $n$-doped material, via enhanced
antiferromagnetic spin correlations, the competition between
antiferromagnetic order and superconductivity.

As far as the modeling of the DF's is concerned the anisotropic dipole
polarizability is modified when compared with the insulator. The dipole
polarization in $x$- and $y$-direction of the Cu and O$_{xy}$ ions in
the CuO plane is assumed to be zero because of the metallic screening
in the plane. Such a modeling already has been applied successfully for
LaCuO in Ref. \onlinecite{24}. Moreover, we have reduced the dipole
polarization of the apex oxygen in $z$-direction in the metallic state.
This would be consistent with a better screening in $c$-direction via
the delocalized Cu$4s$ charge fluctuations, additionally allowed. The
actual values of the dipole polarizability are given in Table
\ref{tabdipol2}.

\begin{table}
\begin{tabular}{c|cccc}
    &  Cu &  O$_{xy}$  &  O$_z$  &  Nd   \\\hline\hline
ab-initio & 8.9 & 7.2 & 8.2 & 12.5 \\\hline
xy  & $-$ &  $-$   & 100$\%$  & 40$\%$\\
z   &40$\%$&  100$\%$  & 25$\%$ & 40$\%$\\
\end{tabular}
\caption{The calculated result for the dipole polarizability according
to the Sternheimer method\cite{30} in units of $a_\text{B}^3$ is given
in the first row for Cu$^{1.22+}$, O$_{xy}^{1.42-}$, O$_z^{1.54-}$ and
Nd$^{2.35+}$ . The anisotropic reduction of the polarizability for the
metal is presented in the last two lines in percent. The $-$ denotes
zero polarizability. 100 percent stands for the ab initio
result.}\label{tabdipol2}
\end{table}

\subsection{PHONON DISPERSION ACROSS THE INSULATOR-METAL TRANSITION}
In this section we propose a modeling of the insulator-metal transition
in NdCuO via the underdoped  phase characterized by the sum rule for
the charge response given in Eq. \eqref{27}. Looking from the
perspective of the insulator this means that in such
localization-delocalization transition expressed by the sum rule the
O$2p$ orbital remains incompressible, insulator-like (localized), while
the Cu$3d$ and Cu$4s$ orbital becomes compressible, metallic upon
$n$-doping.

\begin{figure}
\includegraphics[width=\linewidth]{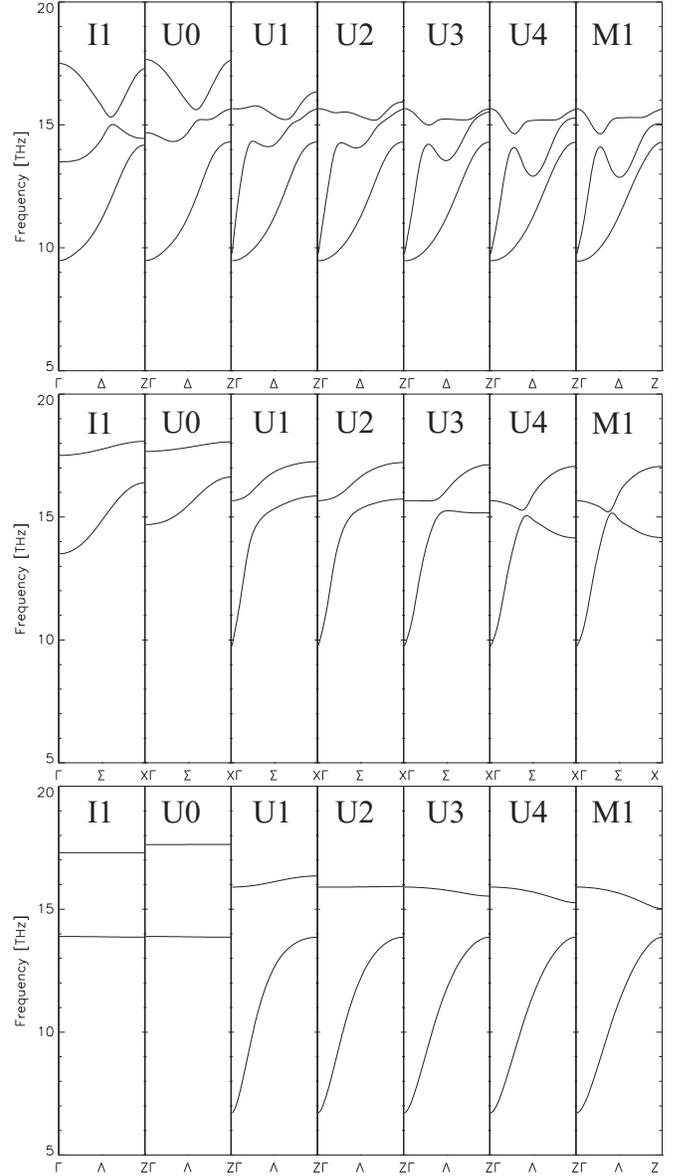}
\caption{Prediction of the anomalous phonon branches of $\Delta_{1}$,
$\Sigma_{1}$ and $\Lambda_{1}$ symmetry, respectively, across the
insulator-metal transition. The different models are denoted from left
to right: I1, U0, U1, U2, U3, U4, M1. I, U, M denote the insulating,
underdoped and metallic state, respectively. The unit of the frequency
is THz. A detailed description of the calculated results in the
different panels is provided in the text.}\label{FIG07}
\end{figure}

In Fig. \ref{FIG07} we present a sequence of model calculations of the
high-energy phonon modes related to the $\Delta_{1}$, $\Sigma_{1}$ and
$\Lambda_{1}$ branches, respectively. The calculations include the
anomalous OBSM i.e. the oxygen half-breathing mode (\dhalbe anomaly),
the planar oxgen breathing mode at the $X$ point (\obx) and the axial
oxygen breathing mode \ozz at the $Z$ point of the BZ, see Sec.
\ref{sec_anticrossing} for the corresponding displacement patterns. In
the leftmost panel the results for the insulator are shown and in the
rightmost panel those for the metal, compare with Fig. \ref{FIG04} and
Fig. \ref{FIG06}, respectively.

For the description of the change of the phonon dispersion across the
insulator-metal transition in a first step we replace the dipole
polarizations of the insulator (model I1) by those of the metal. The
result of this fictitious system is shown in the second panel from the
left (model U0). The system is still an insulator, because the
polarizability related to the CF's has not been changed in the model,
i.e. the Cu$3d$ and O$2p$ orbital is still incompressible according to
the sum rule in Eq. \eqref{24}. The next four panels display our
results for the phonon dispersion across the transition via an
underdoped strange metallic state. The corresponding sequence of models
is denoted U1-U4 in Fig. \ref{FIG07} and the sum rule from Eq.
\eqref{27} is used for the modeling, i.e. the O$2p$ orbital is kept
incompressible. In order to satisfy this requirement some of the matrix
elements of the polarizability must be chosen from the outside because
the system of equations is under-determined. In particular we take the
important $\Pi(\text{Cu$3d$})$ and $\Pi(\text{Cu$4s$})$ component of
the matrix as free parameters in order to calculate their possible
effect on the phonon dispersion when they are varied across the
transition.

The definite variation of these matrix elements for modeling the
sequence U1-U4 can be extracted from Table \ref{table01}. The partial
density of states $\widetilde{Z}_\kappa(\varepsilon_\text{F})$ for the
Cu$3d$ and Cu$4s$ orbital on the right hand side of Eq. \eqref{27} is
taken to be equal to $\Pi(\text{Cu}3d)$ and $\Pi(\text{Cu}4s)$,
respectively. This is approximatively true, because the diagonal
elements of $\Pi_{\kappa\kappa'}$ dominate for the compressible,
metallic components. The matrix elements $\Pi(\text{O}_{xy})$,
$\Pi(\text{Off})$, $\Pi(\text{O}_x-\text{O}_y)$,
$\Pi(\text{Cu}3d-\text{Off})$, $\Pi(\text{O}_{xy}-\text{Off})$ are
taken from the model I1 of the insulator, where the Cu$4s$ orbital is
not involved. Finally, the remaining four unknown parameters, i.e.
$\Pi(\text{Cu}4s-\text{Off})$, $\Pi(\text{Cu}3d-\text{Cu}4s)$,
$\Pi(\text{Cu}4s-\text{O}_{xy})$ and $\Pi(\text{Cu}3d-\text{O}_{xy})$
are determined from the sum rule in Eq. \eqref{27}. Of course such a
procedure is not unique, see in this context also our discussion of a
similar modeling for LaCuO in Ref. \onlinecite{23}. The complete list
of the definite values for the matrix elements of the CF-polarizability
for all the models used in our calculations is summarized in Table
\ref{table01} and \ref{tab4}.

\begin{table}
\begin{tabular}{c|c|c|c|c|c|c}
     &  I1, U0  &   U1  &  U2  &  U3   & U4   &  M1\\ \hline
Cu$3d$ &0.05& 0.05 &   0.20& 1.40   & 2.800  &2.800  \\
Cu$4s$ & $-$  & 0.02 &  0.02& 0.04  & 0.075&0.075 \\
O$2p$  & 0.01 &0.01 &0.01 &0.01 &0.010 &0.030 \\
Off    &  0.02 &  0.02  &  0.02 &  0.02  &  0.020 &  0.020 \\
\end{tabular}
 \caption{Diagonal components $\Pi(\text{Cu$3d$})$, $\Pi(\text{Cu$4s$})$,
 $\Pi(\text{O$2p$})$ and $\Pi(\text{Off})$ of the polarizability matrix $\Pi_{\kappa\kappa'}$
 used in the different models for the calculation of the phonon dispersion of
 NdCuO across the insulator-metal transition via the underdoped state.
 I1 denotes the insulator, M1 the metal and U0-U4 the underdoped state
 simulating different doping levels. Note the different
 dipole polarizations in I1 and U0 as discussed in the text. Units for $\Pi_{\kappa\kappa'}$
 are in $eV^{-1}$.}\label{table01}
\end{table}

\begin{table}
\begin{tabular}{c|c|c|c|}
            &  I1, U0  &   U1, U2, U3, U4   &  M1\\ \hline
O$_x2p$-O$_y2p$ & -0.00100 &-0.00100& $-$\\
Cu$3d$-Off     & -0.00725 & -0.00725 & -0.00250\\
Cu$3d$-O$2p$ & -0.00525&  0.00100 & $-$\\
O$2p$-Off & 0.00225& 0.00225&-0.00250 \\
Cu$3d$-Cu$4s$ & $-$ & 0.02500 & $-$ \\
Cu$4s$-O$2p$ & $-$ & -0.00625 & $-$\\
\end{tabular}
\caption{Values of the off-diagonal elements of the CF-polarizability
in units of $eV^{-1}$ for all the models discussed in the text.
$\Pi(\text{Cu}4s-\text{Off})$ is virtually zero and not given in the
table.}\label{tab4}
\end{table}

The modification of the charge response of the relevant orbitals Cu$3d$
and Cu$4s$ in terms of the polarizability matrix simulating doping in
our approach is given in Table \ref{table01}. We assume that the main
effect introduced by $n$-doping is to populate the hybridized orbitals
of Cu$3d$ and Cu$4s$ character. Thus, the partial density of states
$\widetilde{Z}_{\kappa}(\varepsilon_{F})$ for these orbitals should
increase and correspondingly the components $\Pi(\text{Cu$3d$})$ and
$\Pi(\text{Cu$4s$})$ of the polarizability. In Table \ref{table01} we
specify the values for $\Pi(\text{Cu$3d$})$ and $\Pi(\text{Cu$4s$})$
which have been used for the calculation of the phonon dispersion in
the models U1-U4 as shown in Fig. \ref{FIG07}. To act as a reference
the parameters for the insulating and metallic state, I1 and M1, are
also given. Increasing $\Pi(\text{Cu$3d$})$ and $\Pi(\text{Cu$4s$})$ in
the sequence U1-U4 simulates an increase of $n$-doping in our modeling.
This is carried out until the magnitude of these quantities are conform
with model M1 for the metallic phase. In the latter, the formerly
incompressible O$2p$ orbital is allowed to become compressible and now
additionally may contribute to the metallic properties in NdCuO. At the
end, we obtain the phonon dispersion for the model M1 shown in the
rightmost panel where all orbital degrees of freedom are compressible,
metallic. A good agreement with the experimental dispersion curves
including all their unusual features is achieved.

Concerning the change of the electronic properties across the
insulator-metal transition via the strange metallic state, which should
display non-Fermi-liquid properties according to our modeling\cite{21},
the main effect is related to an abrupt change of the electronic
structure near the Fermi level and is globally expressed by an increase
with doping of the partial density of states
$\widetilde{Z}_\kappa(\varepsilon_\text{F})$ for the compressible,
metallic Cu$3d$ and Cu$4s$ orbital, respectively. The sudden loss in
the DOS at the Fermi level (pseudogap) when passing from the model of
the normal Fermi liquid with a large FS to a model for the strange
metallic state with some reconstructed smaller FS by an orbital
resolved compressibility-incompressibility transition can be expected
to show up in all physical properties which scale with the DOS at the
Fermi level, like thermodynamic properties, e.g. the electronic
specific heat coefficient, transport properties as the resistivity or
charge- and spin correlation functions. In context with our generic
modeling of the strange metallic state in the cuprates and the related
FS reconstruction, most likely conspiring with orbital-dependent band
distortions, it is interesting to note that quite recently the
existence of a small FS has been approved experimentally by the
observation of quantum oscillations of the Hall resistance in an
underdoped cuprate, see Ref. \onlinecite{48} and also Ref.
\onlinecite{49}. Finally, such a doping dependent FS reconstruction
could be a means to explain, at least partly, the doping dependence of
the unusual normal-state properties of the cuprates.

While the pseudogap in the normal state of the cuprates is due to the
incompressible, insulator-like states in our modeling, a distinct gap
below T$_c$ related to superconductivity can result from pairing of the
doped charge carriers at a lower energy scale in the compressible,
metallic states. Interestingly, according to the sum rules in Eqs.
\eqref{26}, \eqref{27} the condition for incompressibility and
compressibility are not independent because of the off-diagonal
elements of the polarizability matrix. Thus, it can be expected that
below T$_c$ the pseudogap and the superconducting gap are correlated
and above T$_c$ the small FS (arcs or pockets) and the pseudogap are
related too. For example, a negative value of the
(Cu$3d$-O$_\text{xy}$) matrix element necessary to make the Cu$3d$
state incompressible according to Eq. \eqref{26}, as required for
underdoping, simultaneously will decrease the PDOS for the compressible
O$2p$ state. This correlates with a decrease of the small FS in the
underdoped normal state and predicts a reduction of the superconducting
gap if a BCS scenario is assumed. On the other hand, the matrix
elements of the polarizability for Cu$3d$ and O$_\text{xy}$ decrease
when doping is reduced towards the insulating phase. So, the pseudogap
as a loss in the density of states at the Fermi energy will increase
with less doping.

As already mentioned, when going from the underdoped state via the
optimally to the overdoped state in our modeling \textit{all} the
orbitals become compressible, metallic. The pseudopgap effect related
to the localized, incompressible Cu$3d$ orbital in underdoped $p$-type
cuprates vanishes and the compressibility coming along with a
delocalization of all states corresponds to a recovering of the
reconstructed small FS to a large one consistent with ARPES
experiments. The formerly incompressible Cu$3d$ orbital now dominates
the DOS at $\varepsilon_\text{F}$ in the modeling. Additionally, a
(virtual) Cu$4s$ contribution develops at $\varepsilon_\text{F}$, which
has shown its important fingerprints in the forming of the OBSM phonon
anomalies upon doping. Altogether, a \textit{critical point} which
separates two qualitatively different electronic ground states is
reached in our modeling when the incompressible orbitals become
compressible accompanied by enhanced fermionic particle-number
fluctuations, a transformation of the FS and the vanishing of the
pseudogap effect. The orbital resolved compressibility of Cu$3d$ in
$p$-type cuprates and of O$2p$ in $n$-type cuprates is a certain
measure of the local particle number fluctuations in the corresponding
orbitals. The incompressibility of these orbitals in the underdoped
state corresponds only approximatively to locally conserved particle
numbers, because of the finite interaction strength in a realistic
approach. In case, such quantities would be strictly conserved the
latter are linked by general principles with \textit{local gauge
symmetry} of the theory. Using reduced model Hamiltonians (like the
infinite U Anderson or Hubbard model) the gauge symmetry approach has
been extensively discussed in the literature in context with the
cuprates, see. e.g. Ref. \onlinecite{50}. Physically, the partial
incompressibility expressed by
$\widetilde{Z}_\text{Cu$3d$}(\varepsilon_{F})=0$ for $p$-doped and
$\widetilde{Z}_\text{O$2p$}(\varepsilon_{F})=0$ for $n$-doped materials
in our modeling and thus the corresponding reduced local particle
density fluctuations are mainly related to localization by strong
correlations and localization effects of ionic origin, respectively.
This constrains the low-energy dynamics of the electrons in the
underdoped state, is responsible for the destruction of the large FS of
the Fermi-liquid and is responsible for the non-Fermi liquid behaviour,
correlating with a reconstructed FS.

A superconducting state may become possible through an attractive
interaction leading to pairing and destabilization of the FS as a
result of an interplay of spin, charge and lattice degrees of
freedom\cite{19}. Spin-fluctuations, phonons and in the case of the
optimally to overdoped state also coherent CF's along the $c$-axis in
form of low lying phonon-plasmon modes provide a \textit{retarded}
contribution to the pairing interaction, while CF's and residual
anti-ferromagnetic spin-fluctuations in the Cu-O-plane contribute a
\textit{non retarded}, approximatively instantaneous, part to the
interaction at a higher energy scale.

The large FS in the optimally to overdoped state also gives rise to a
corresponding larger phase space for pairing as compared with the
underdoped phase, in case some small FS due to the localization of the
Cu$3d$ or O$2p$ in $p$- or $n$-doped cuprates, respectively, is
representative for this state. According to the additional
compressibility of the Cu$3d$ orbital or O$2p$, respectively, particle
number fluctuations are enhanced as compared with the underdoped,
partial incompressible state. Thus, phase coherence is established more
easily according to the particle-number-phase uncertainity relationship
$\Delta N\,\Delta\phi \geq 1$. A similar duality also exists locally,
Ref. \onlinecite{51}. On the other hand, the reduction of
particle-number-fluctuations in the underdoped state may clear the way
for the formation of electron pairs at a higher energy scale well above
the onset of phase coherence at T$_c$. Such a two-particle contribution
of preformed pairs would also be accompanied by a reduction in the DOS
which would add to the loss in the DOS from the orbital selective
incompressibility in the single-particle channel.

Furthermore, it should be remarked that within our model for the
$p$-doped cuprates according to \eqref{26} the compressible, metallic
O$2p$ component of the electronic state is always connected to
superconductivity, while the Cu$3d$ and Cu$4s$ component additional
comes into play for optimal to overdoping. On the other hand after the
sum rule from \eqref{27} the situation is reversed for the $n$-doped
case, i.e. superconductivity in the underdoped phase is orbital
selective in the spirit of our modeling.

The incompressible, localized Cu$3d$ orbital in $p$-doped and the O$2p$
orbital in $n$-doped cuprates also can be expected to promote
antiferromagnetic spin fluctuations.

An ab initio calculation of the polarizability matrix $\Pi$ for the
HTSC's, say beyond LDA, which should include the strong electron
correlations and also the strong nonlocal EPI effects in the self
energy of the electrons, seems not possible in the near future. In
particular this holds for the underdoped state of the cuprates. Thus,
the PDOS, $Z_\kappa(\varepsilon_\text{F})$,
$\widetilde{Z}_\kappa(\varepsilon_\text{F})$ and the matrix elements of
$\Pi$ appearing in the various sum rules \eqref{22}-\eqref{27} can be
understood as quantities renormalized by these interactions. Their
definite values are presently unknown. A LDA result possibly may be a
guide for the optimally to overdoped phase. In general, the pseudogap
and the superconducting gap and the correlation in our approach is of
course affected by these renormalization effects. In any case, in our
modeling the matrix elements of $\Pi$ and the PDOS can be treated as
microscopically well defined parameters of the theory constrained by
the rigorous compressibility sum rules.

Finally, the strong doping dependent nonlocal EPI effects accompanied
by electronic CF's as found for the OBSM and, in particular, also for
the nonadiabatic phonons of polaronic character (mixed phonon-plasmon
modes) propagating in a small region around the $c$-axis of the
HTSC's\cite{19} are certainly important to understand the unusual
doping dependent \textit{isotope effect} in the cuprates, because of
their contribution to the electronic self energy.

As already mentioned earlier, $\Pi(\text{O$2p$})$ in model M1 is much
smaller as for the modeling of the optimally doped metallic state of
$p$-doped LaCuO because of a stronger localization of O$2p$ in NdCuO.
In case of LaCuO we have obtained $\Pi(\text{O$2p$})=0.2 eV^{-1}$, Ref.
\onlinecite{23}, which has to be compared with $\Pi(\text{O$2p$})=0.03
eV^{-1}$ in case of NdCuO (Table \ref{table01}). This small value
points to a much weaker pseudogap effect in a n-type material as NdCuO
as compared with the $p$-doped cuprates.

As remarked in the introduction, in the ARPES experiments a large
LDA-like FS of hole type evolves at optimal doping in metallic
(superconducting) NdCuO, similar as in $p$-doped LaCuO, from the
electron pockets seen in non-superconducting underdoped NdCuO. This
suggests that holes might play a similar role in both types of
superconductors. Moreover, as discussed in Sec. \ref{sec_NCO_metall}
the CF-polarizability of the delocalized Cu$4s$ orbital becomes
important in the metallic phase pointing to a substantial mixing of
Cu$4s$ with Cu$3d$ and O$2p$ in the wavefunction at the Fermi level.
Figuratively, this may be interpreted by the emergence of holes in the
O$2p$ states which creates room for a delocalization of the Cu related
states in particular via the Cu$4s$ orbital. The effect of the
insulator-(strange)-metal transition has as a consequence a qualitative
renormalization of the phonon modes as can be extracted by comparing
the results from the second and third panel of Fig. \ref{FIG07}. These
qualitative changes are due to the closing of the LO-TO splittings of
the E$_{u}$ modes along the $\Delta$ and $\Sigma$ direction at the
$\Gamma$ point and the vanishing of the A$_{2u}$ discontinuities along
the $\Lambda$-direction if an adiabatic charge response along the
$c$-axis is assumed as in our modeling.

From our calculations of the electronic charge response and the
resulting phonon dynamics the following qualitative physical picture is
consistent with our modeling of the electronic state in the cuprates,
see also the detailed discussion for the $p$-doped case in Ref.
\onlinecite{21}. In the underdoped $p$-type material metallic hole
carriers are related according to the sum rule from Eq. \eqref{26} to
the compressible O$2p$ orbitals in the CuO plane. Cu-sites are avoided
by the holes because of the strong on-site Coulomb interaction,
U$_{3d}$, of the  Cu$3d$ orbitals in the cuprates which are treated as
incompressible, insulatorlike. Thus, there will be no repulsive core of
the Coulomb interaction at the Cu-site for the holes in the
superconducting state. This is compatible for example with $d$-wave
superconductivity which on his part is favoured by a specific feature
of the cuprates, namely, their common  CuO square lattice.

In the optimally $p$-doped state the Cu$3d$/$4s$ orbitals also become
compressible, metallic in our modeling because doping of more holes in
the CuO plane creates room for a delocalization and a corresponding
gain in kinetic energy for the Cu related states, in particular for
Cu$4s$ by hybrization with the O$2p$ orbitals. The mixing in of the
Cu$4s$ state also may contribute a weak $s$-component for pairing.
Moreover, the loss of density of states at $\varepsilon_{F}$
(pseudogap) expressed  by Eq. \eqref{26}, due to the incompressible Cu
states in the underdoped material, is restored upon doping and so these
states additionally become available for pairing.

In underdoped $n$-type materials Cu$3d$/$4s$ orbitals are allowed to be
compressible according to Eq. \eqref{27} while O$2p$ remains
incompressible. This means that we have an electron dominated transport
in the normal state and a repulsive core of the Coulomb interaction at
the Cu-site which is unfavourable for $s$-wave pairing of electrons
because of the large on-site Coulomb interaction U$_{3d}$, while holes
in compressible, metallic O$2p$ orbitals are not present unlike to the
case in the underdoped state of the $p$-doped materials, see Eq.
\eqref{26}. However, $s$-wave pairing could become a possibility with
increased doping because of the small on-site Coulomb interaction
U$_{4s}$. Note in this context, that the importance of the Cu$4s$
orbital for the charge response across the insulator-metal transition
via the underdoped state is growing in the modeling of the phonon
dispersion in Fig. \ref{FIG07}. $D$-wave pairing via hole carriers is
not allowed in the modeling for the underdoped state because the O$2p$
orbital is incompressible. On the other hand, in the optimally doped
case the latter becomes compressible, metallic too and hole carriers in
the O$2p$ orbitals can emerge like in $p$-doped materials. Remember
that the FS of optimally doped NdCuO is of the hole-type and LDA-like,
Refs. \onlinecite{25,26}, similar as for optimal $p$-doped LaCuO. In
this way the repulsive core at Cu of the Coulomb interaction is avoided
for the holes. Altogether, this leads to the conclusion that holes in
the O$2p$ states should play a similar role for $d$-wave pairing in
\textit{both} types of superconductors. A conclusion that holes are
responsible for the superconductivity also in $n$-doped cuprates quite
recently has been drawn for the electron-doped superconductor
Pr$_{2-x}$Ce$_x$CuO$_4$ from the experimental side, i.e. by
measurements of the resistivity and Hall angle as a function of doping
and temperature, Ref. \onlinecite{52}.

Nonadiabatic modifications to investigate the charge response around
the "ionic" $c$-axis have been employed in Refs. \onlinecite{19,38}.
Given the nearly two-dimensional electronic structure of the cuprates
and thus a very weak interlayer coupling electron dynamics and phonon
dynamics will be on the same time scale in a small region around the
$c$-axis. This needs a \textit{nonadiabatic} treatment with dynamical
screening of the bare long-ranged Coulomb interaction and leads to a
strong nonlocal, nonadiabatic EPI of \textit{polaronic} character with
phonon-plasmon mixing in the metallic state of the HTSC's. Such a
strong coupling of the phonons to the electrons along the $c$-axis acts
against a coherent interlayer hopping along this axis, helps on a
confinement of the electrons in the CuO plane and may trigger a
metallic to a nonmetallic crossover in the $c$-axis resistivity in case
of a sufficiently low lying $c$-axis plasmon. The energetic position of
the plasmon is of course material specific and depends on the
anisotropy and doping level of the compound, Refs. \onlinecite{19,38}.

Following the dispersion of the $\Delta_{1}$ modes in the metallic
phase shown in the panels to the right of the second panel we find a
continuous variation of the curves. In particular for the $\Delta_{1}$
branch with the second highest frequencies we detect the development of
a local minimum that can be identified with the \dhalbe anomaly found
in the optimally doped metallic phase of NdCuO by INS. Also the
$\Delta_{1}$ branch with the highest frequency develops a local minimum
in full agreement with the experiments. The evolution to the two
highest $\Sigma_{1}$ branches in Fig. \ref{FIG07} also converges to the
experimental results for the optimally doped material, compare with
Fig. \ref{FIG06}. From these calculations it follows that a continuous
enhancement of the polarizability related to CF's of the Cu$3d$ and in
particular Cu$4s$ orbital in the underdoped strange metallic state
leads to the correct result for the optimally doped case.

\section{PHONONANOMALIES AND ANTICROSSING BEHAVIOUR}\label{sec_anticrossing}
In the electron-doped material NdCuO the dispersion of the branches
with high frequency comprising the anomalous OBSM is more complex than
in the hole-doped LaCuO. This becomes obvious when contrasting the
results displayed in Fig. \ref{FIG04} and Fig. \ref{FIG06} with the
analogous results for LaCuO, see e.g. Refs. \onlinecite{19,21,24}. The
reason is that in NdCuO there is another branch of the same
$\Delta_{1}$-symmetry which interacts with the branch containing the
OBSM. Moreover, in the metallic phase (Fig. \ref{FIG06}) there is
additionally a third branch of the same symmetry in the $\Delta$
direction close-by in frequency. Thus, we will have anticrossing
phenomena being absent in LaCuO because of the different crystal
structure. Note, that a different lattice structure for LaCuO and NdCuO
is also not in favour of an electron-hole symmetry.

\begin{figure}%
\begin{minipage}{0.5\linewidth}%
\includegraphics[width=\linewidth]{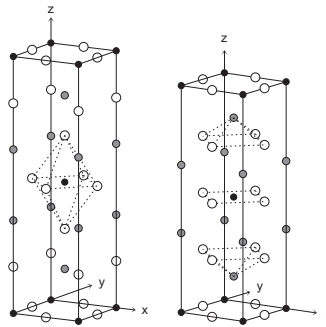}%
\end{minipage}%
\hspace{0.1\linewidth}%
\begin{minipage}{0.4\linewidth}%
\caption{Elementary cell of tetragonal LaCuO ($T$-structure, left) and
NdCuO ($T'$-structure, right). $\bullet$: Cu, $\circ$: O,
\textcolor[rgb]{0.50,0.50,0.50}{$\bullet$}: La, Nd}\label{FIG08}%
\end{minipage}%
\end{figure}

\begin{figure*}
\includegraphics[width=0.7\linewidth]{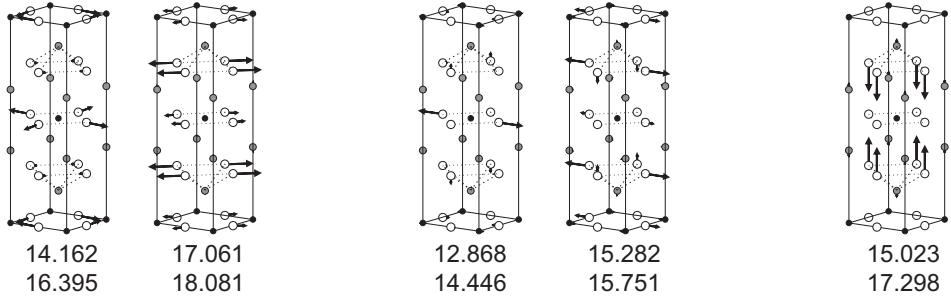}
\caption{Displacement patterns of the anomalous OBSM \obx (left),
\dhalbe (middle), \ozz (right) for NdCuO. The calculated frequencies in
units of THz for the metallic phase are listed in the first row and
those for the insulating phase in the second row below the
patterns.}\label{FIG09}
\end{figure*}

From the perspective of LaCuO (T-structure) the apex oxygen is shifted
to a new position in NdCuO (T'-structure), see Fig. \ref{FIG08}. Here the bonding
environment for the oxygen has changed as compared to LaCuO and
vibrations with higher energies of the oxygen ions parallel to the CuO
plane in the NdO "layer" become possible (Fig. \ref{FIG09}) which may interact
with the OBSM. In this figure also the calculated frequencies of the
anomalous OBSM are listed for both, the metallic and the insulating
phase. From these data and from Fig. \ref{FIG04} and Fig. \ref{FIG06} an anomalous
softening of the OBSM across the insulator-metal transition can be
extracted, similar to our findings for the $p$-doped cuprates. This,
ultimately supports the generic nature of the phonon anomalies in
$p$-doped and $n$-doped HTSC's.

It is quite instructive to compare the calculated renormalization of
the OBSM due to strong nonlocal EPI of CF-type across the
insulator-metal transition in NdCuO and LaCuO, respectively.
Quantitatively we obtain for \obx, \dhalbe and \ozz the following
softening in THz (NdCuO first number, LaCuO second number). \obx:
2.233, 2.539; \dhalbe: 1.578,\,1.452; \ozz: 2.275,\,2.962. While the in
plane polarized OBSM show a similar renormalization, the softening for
\ozz is stronger and clearly larger in LaCuO. As far as the anomalous
large linewidth of the \ozz mode found in the experiments Refs.
\onlinecite{3,9,18} is concerned we refer to our discussion of this
mode in a \textit{nonadiabatic} phonon-plasmon scenario \cite{19,38}.

Importantly, strongly coupling modes, like \obx and \ozz, which involve
momentum transfer between anti-nodal regions of the Fermi-surface
should contribute via the strong nonlocal EPI to electron self-energy
corrections in the $(\pi,0,0)$-region and in particular to the
anti-nodal pseudogap in the normal state. Moreover, there is an
enhanced contribution to the self-energy in the superconducting state
because of the density of state enhancement in these regions due to the
opening of a gap with $d$-wave symmetry, Ref. \onlinecite{53}. While in
a $d$-wave superconductor \obx connects regions with different sign of
the order parameter, \ozz connects regions with the same sign. The
\dhalbe anomaly on the other hand involves momentum transfer between
nodal regions and could be important to understand the corresponding
self-energy corrections, in particular the nodal kink observed in ARPES
experiments, Refs. \onlinecite{1,2}.

Inspection of the displacement patterns in Fig. \ref{FIG09}
demonstrates that the anomalous OBSM \obx, \dhalbe defined  by oxygen
breathing vibrations in the CuO plane always exhibit the lower
frequency of the two modes with the same symmetry. The modes with the
higher frequencies are essentially confined in the NdO "layer". In the
axial oxygen breathing mode \ozz at the $Z$ point, corresponding to the
apex oxygen breathing mode in LaCuO, all the out of plane oxygens
vibrate in phase against the CuO plane. Therefore, this mode may induce
in the metallic state strong CF's between the CuO planes (interplane
charge transfer), similar as in the case of LaCuO, see e.g. Ref.
\onlinecite{21} and Sec. \ref{sec_chargeresponse}.

\begin{figure}
\includegraphics[height=\linewidth,angle=90]{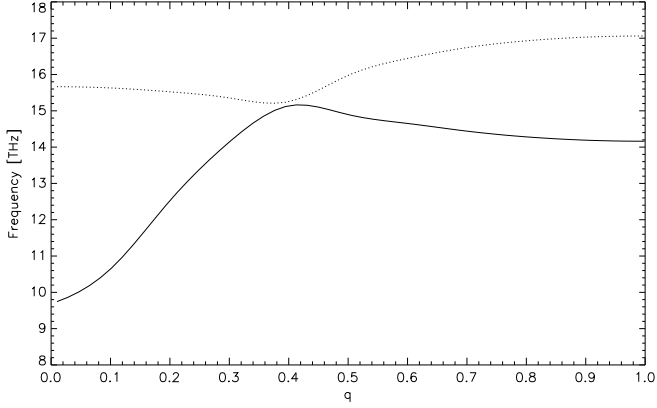}
\caption{Phonon dispersion of the two interacting $\Sigma_{1}\sim
\frac{\pi}{a}q(1,1,0)$ branches from Fig. \ref{FIG06} displayed with a
higher resolution in order to study the anticrossing
effect.}\label{FIG10}
\end{figure}

\begin{figure*}
\includegraphics[width=0.7\linewidth]{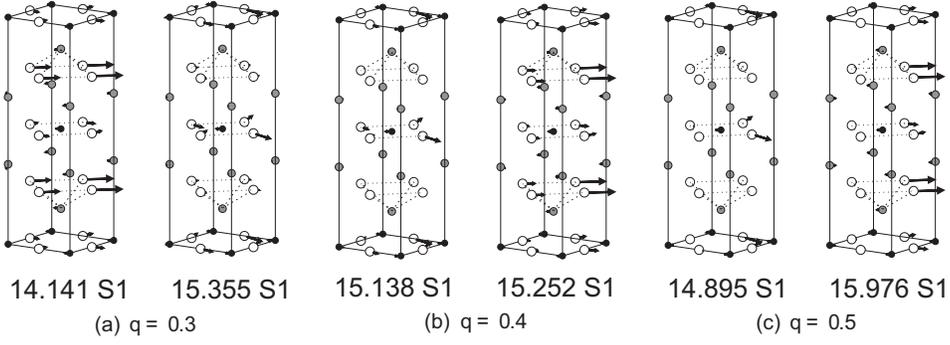}
\caption{Displacement pattern of the two interacting $\Sigma_1$-modes
from Fig. \ref{FIG10} at $q=0.3$, $q=0.4$ and $q=0.5$ in the region
around the anticrossing point localized at about $q=0.4$. The
frequencies of the corresponding modes given below the patterns are in
units of THz.}\label{FIG11}
\end{figure*}

Now, let us study the anticrossing effects of the high-energy modes in
NdCuO incorporating the anomalous OBSM in the $\Delta$ and $\Sigma$
direction. We restrict our discussion to the metallic state. In Fig.
\ref{FIG10} the two interacting $\Sigma_{1}$ branches are displayed
with a higher resolution than in Fig. \ref{FIG06}. From Fig.
\ref{FIG10} the anticrossing point can be localized at about $q = 0.4$.
This may be approved by the calculation of the displacement patterns of
the corresponding eigenvectors. Typical for anticrossing the latter
have changed their character (Fig. \ref{FIG11}) when going from $q =
0.3$ to $q = 0.5$.

\begin{figure}
\includegraphics[height=\linewidth,angle=90]{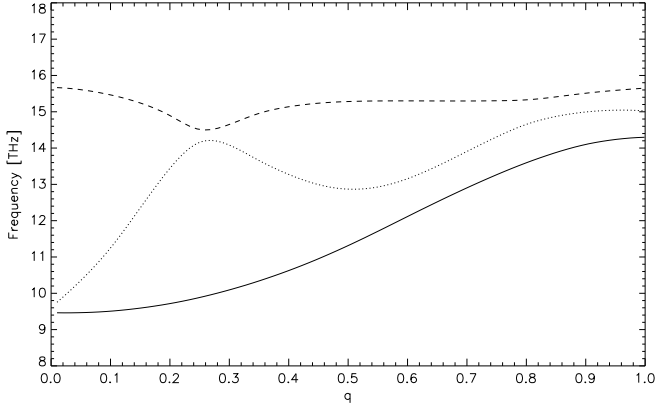}
\caption{Phonon dispersion of the three interacting $\Delta_{1}\sim
(1,0,0)$ branches from Fig. \ref{FIG06} shown with a higher resolution
in order to study normal anticrossing in the sector between $q = 0.2$
and $q = 0.3$ and complex anticrossing in the sector between $q = 0.7$
and $q = 1.0$. $q=0.0$ represents the $\Gamma$ point and $q=1.0$ means
the $Z=\frac{2\pi}{c}(0,0,1)$ point in the BZ.}\label{FIG12}
\end{figure}

\begin{figure*}
\includegraphics[width=0.8\linewidth]{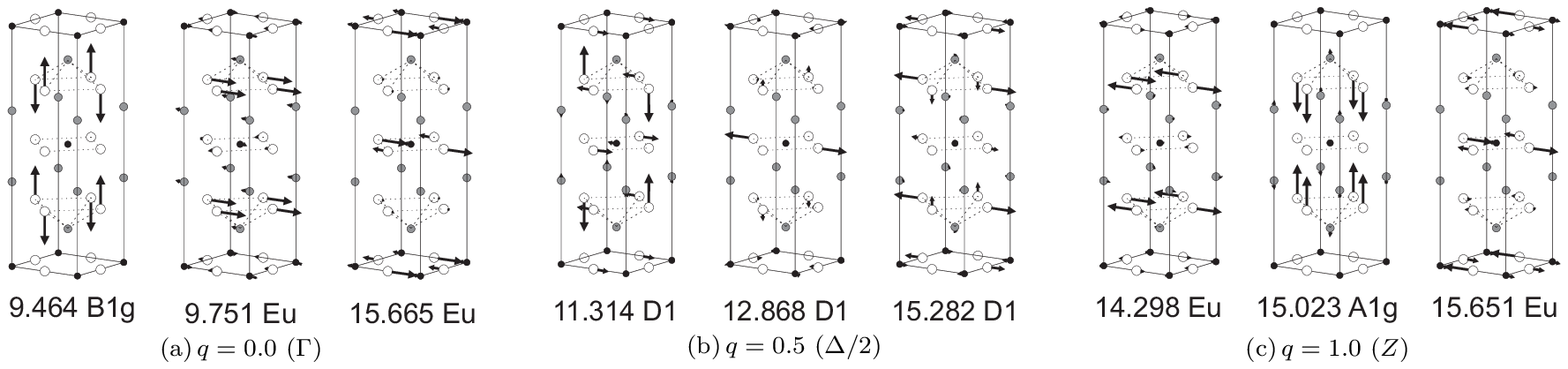}
\caption{Displacement pattern of the three interacting $\Delta_1$-modes
from Fig. \ref{FIG12} at $q = 0$ ($\Gamma$ point), $q = 0.5$
($\Delta/2$) and $q = 1$ ($Z$ point). The frequencies of the
corresponding modes given below the patterns are in units of
THz.}\label{FIG13}
\end{figure*}

The situation for the three interacting high-frequency branches of
$\Delta_{1}$ symmetry is more complex. These branches are shown with a
higher resolution in Fig. \ref{FIG12}. From our calculations we find an
anticrossing of the two highest $\Delta_{1}$ branches between $q = 0.2$
and $q = 0.3$. In the sector between $q = 0.7$ and $q = 1.0$ ($Z$
point) we find a complex anticrossing scenario between three
interacting $\Delta_{1}$ branches. The highest mode at $q = 0.7$ ends
as the lowest mode at the $Z$ point. The lowest mode at $q = 0.7$
becomes the \ozz mode with the second highest frequency at the $Z$
point, compare with Fig. \ref{FIG13}, where the displacement patterns
of the relevant modes are given for the $\Gamma$ point $(q = 0)$,
$\Delta/2$ ($q = 0.5$) and the $Z$ point ($q = 1.0$), respectively.
Finally, the mode with the second highest frequency at $q = 0.7$ winds
up to the highest frequency at $Z$.

\section{PHONON-INDUCED CHARGE RESPONSE OF THE OBSM}\label{sec_chargeresponse}

\begin{figure*}%
\begin{minipage}{0.30\linewidth}%
  \centering%
  \subfigure[\dhalbe]{\includegraphics[width=\linewidth]{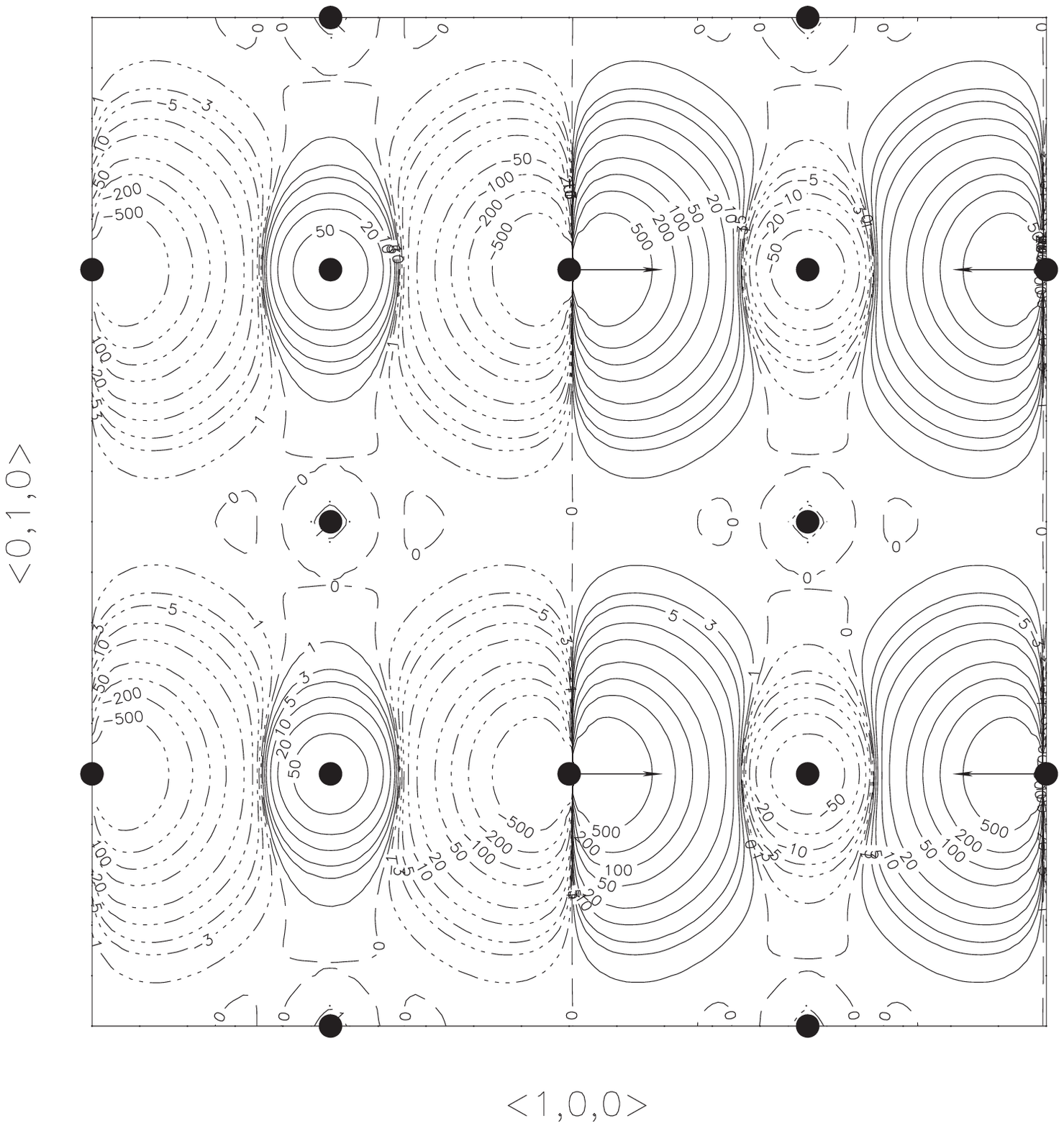}}%
\end{minipage}%
\hspace{0.04\linewidth}%
\begin{minipage}{0.30\linewidth}%
  \centering%
  \subfigure[\dhalbe]{\includegraphics[width=\linewidth]{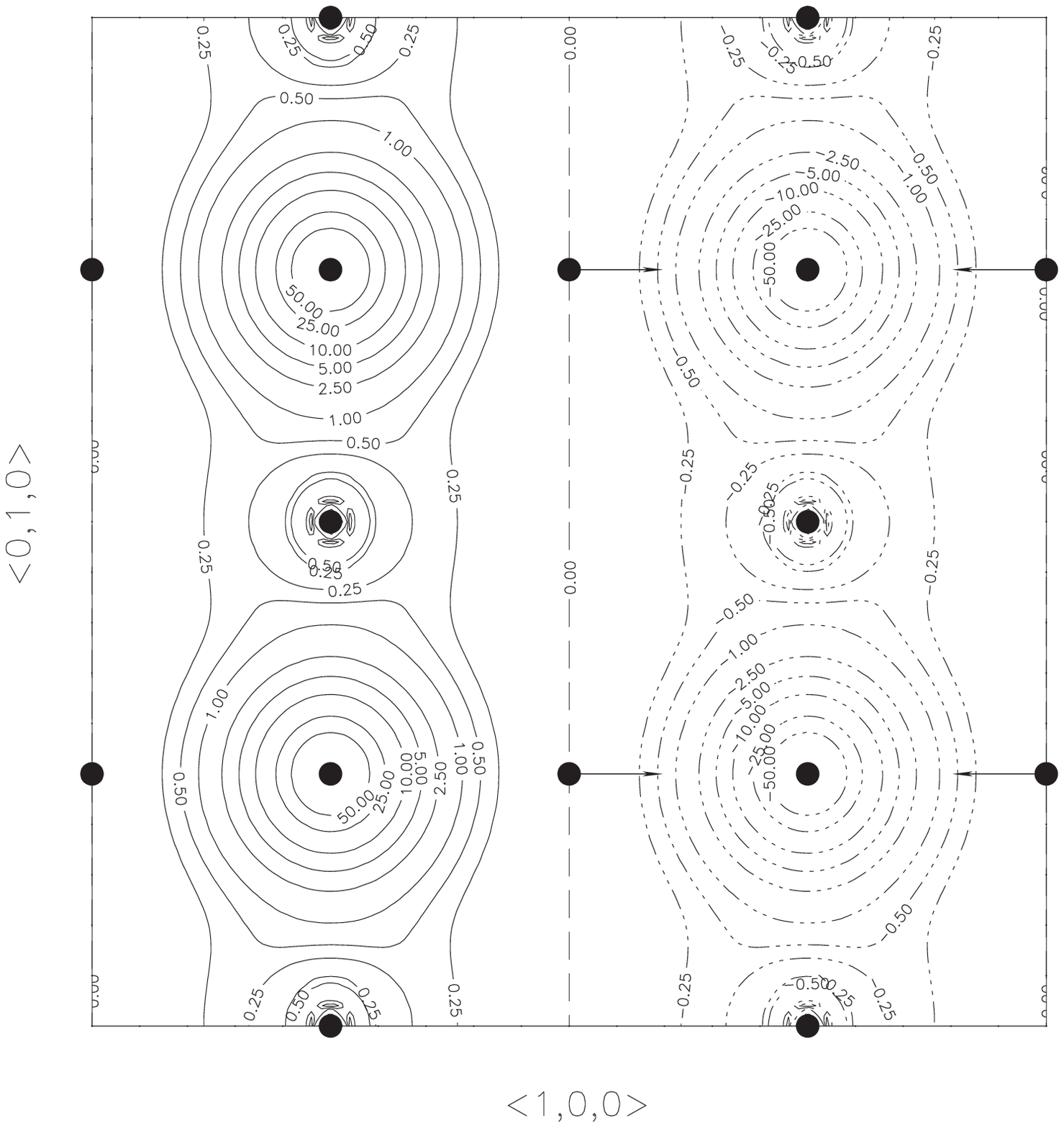}}%
\end{minipage}%
\hspace{0.04\linewidth}%
\begin{minipage}{0.30\linewidth}%
  \centering%
  \subfigure[\ozz]{\includegraphics[width=\linewidth]{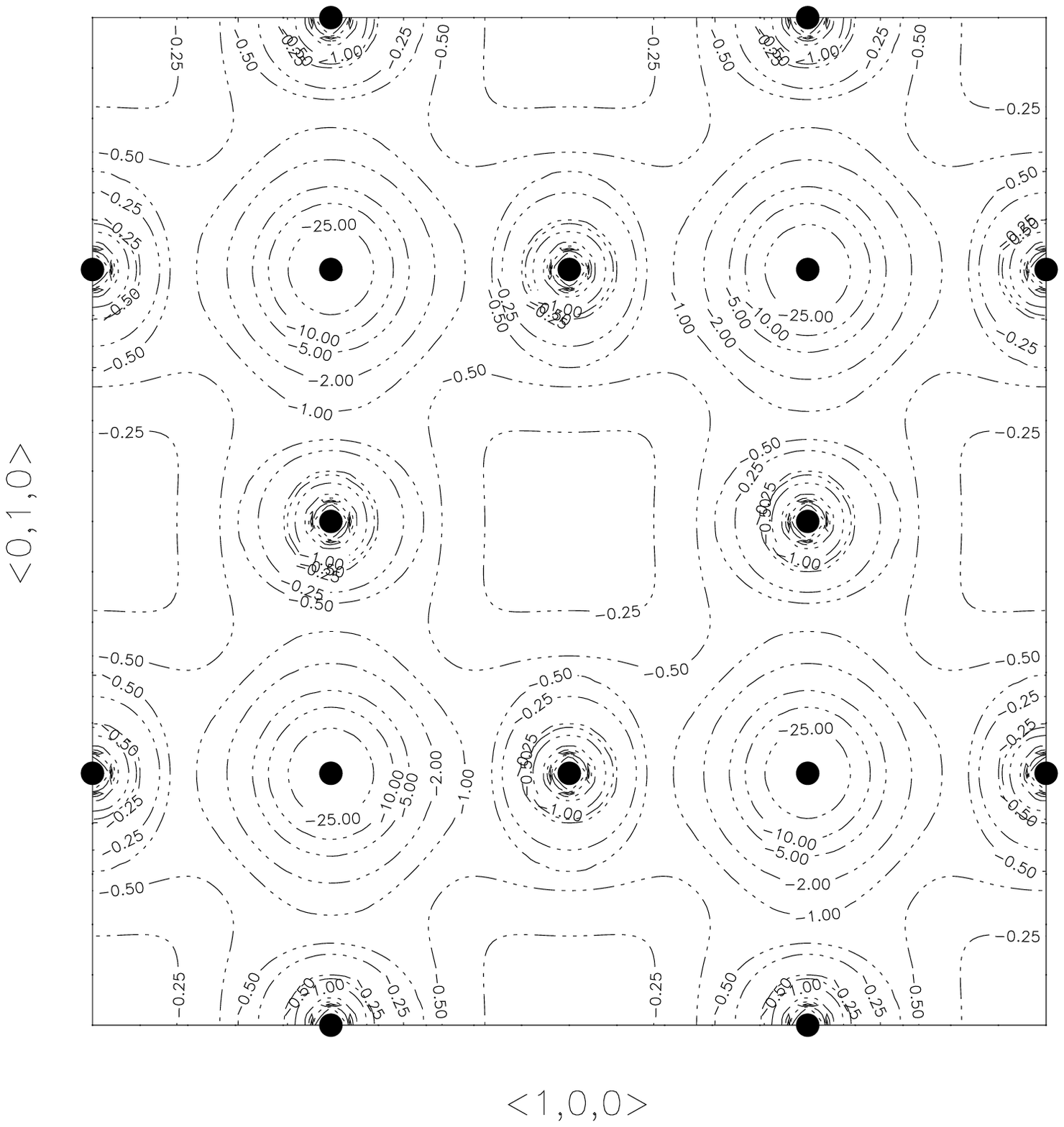}}%
\end{minipage}%

\begin{minipage}{0.30\linewidth}%
  \centering%
  \subfigure[\ozz]{\includegraphics[width=0.4\linewidth]{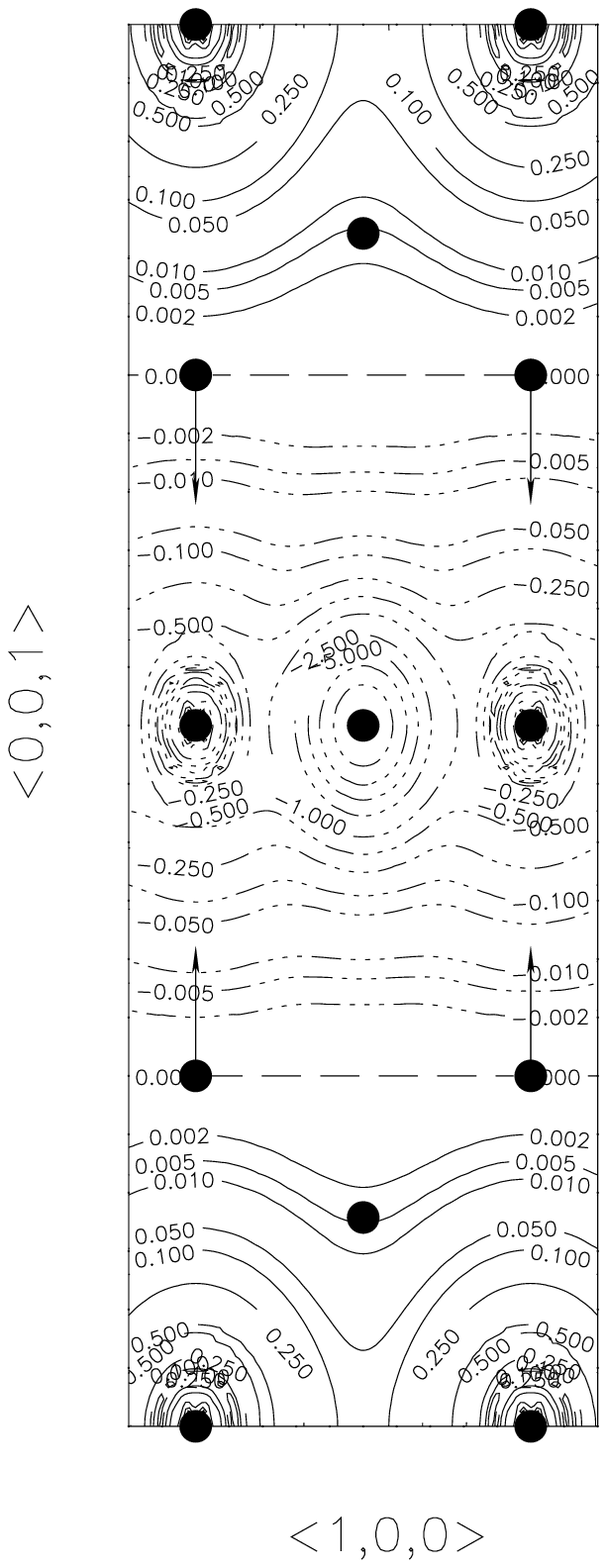}}%
\end{minipage}%
\begin{minipage}{0.30\linewidth}%
  \centering%
  \subfigure[\obx]{\includegraphics[width=\linewidth]{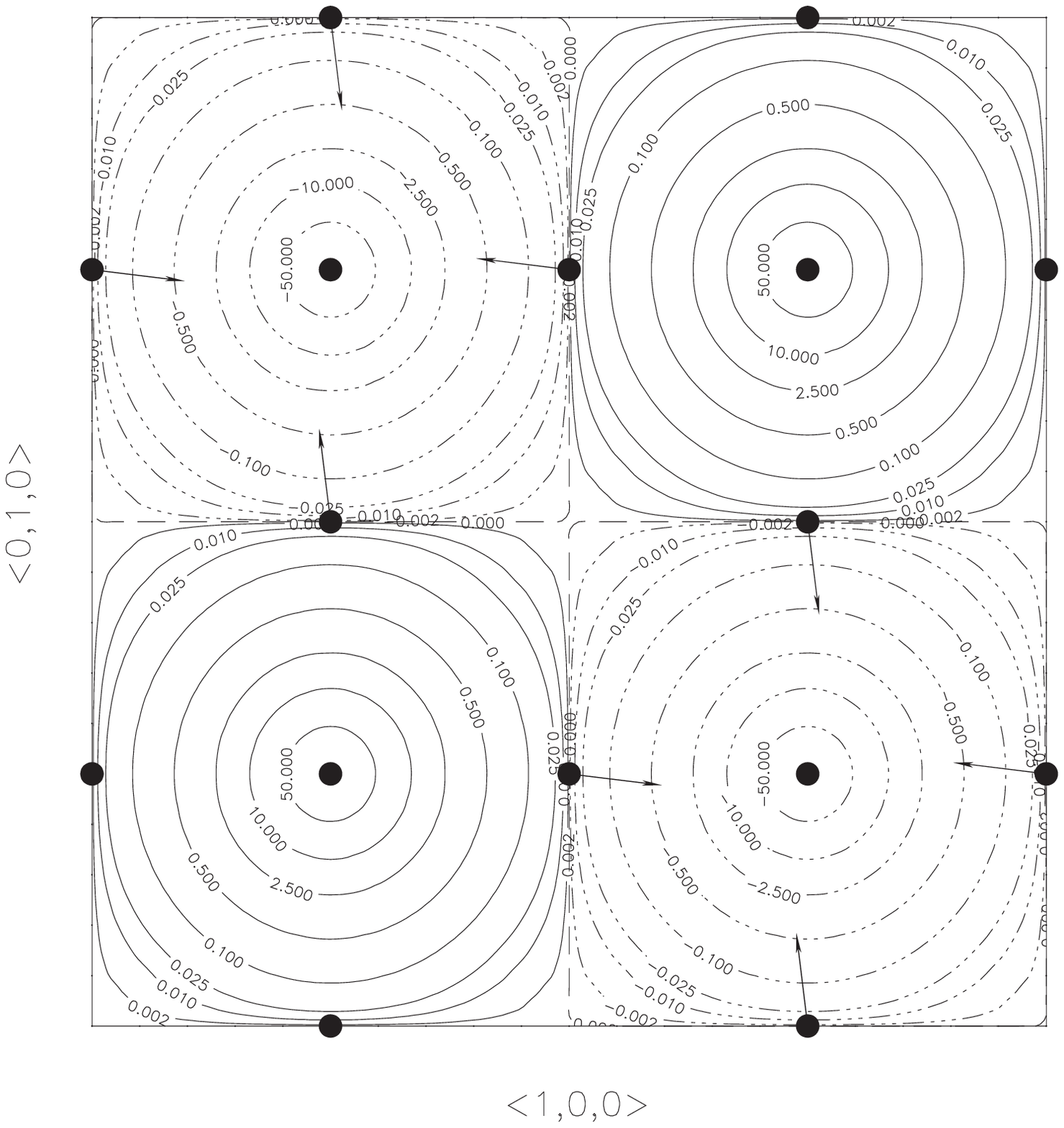}}%
\end{minipage}%
\hspace{0.09\linewidth}%
\begin{minipage}{0.30\linewidth}%
  \caption{Contour plots of the phonon-induced charge
density redistribution $\delta\rho$, according to Eqs. \eqref{28},
\eqref{29} for the OBSM in metallic NdCuO. The units are
$10^{-4}/a_\text{B}^{3}$. Full lines ($-\!\!\!-$) mean that electrons
are accumulated in the corresponding region of space and broken lines
(- $\cdots$) represent regions where the electrons are pushed away. The
arrows denote the  displacement of the oxygen ions (a) total charge
redistribution, i.e. $\delta\rho_\text{r} + \delta\rho_\text{n}$ for
the \dhalbe anomaly (b) nonlocal part $\delta\rho_\text{n}$ of the
charge redistribution for the \dhalbe anomaly. (c), (d)
$\delta\rho_{n}$ for \ozz. (e)
$\delta\rho_{n}$ for \obx.}\label{FIG14}%
\end{minipage}%
\end{figure*}%

In Fig. \ref{FIG14} the result of the calculations of the
phonon-induced redistribution of the charge density $\delta\rho$ is
given for the anomalous OBSM in the metallic phase of NdCuO. Part (a)
displays for the $\Delta_{1}/2$ anomaly the total redistribution
including, both, the local, rigid contribution as obtained from the
local EPI defined by the RIM in Sec. \ref{sec_dispersions} as well as
the nonlocal, nonrigid part of the charge response related to the
nonlocal EPI effects mediated by the CF's. Part (b) gives exclusively
nonlocal contributions of $\delta\rho$, i.e.
\begin{equation}
\delta\rho_\text{n}(\vc{r},\vc{q}\sigma) = \sum\limits_{\vc{a},\kappa}
\delta\zeta^\vc{a}_{\kappa}(\vc{q}\sigma)\rho_{\kappa}(\vc{r} -
\vc{R}^\vc{a}_{\kappa}), \label{28}
\end{equation}
with the CF's according to Eq. \eqref{18} and the form factors
$\rho_{\kappa}$ for the CF's as in Eq. \eqref{1}. The total
phonon-induced charge redistribution can be obtained by adding to Eq.
\eqref{28} the rigidly displayed unperturbed densities, i.e.
\begin{align}\nonumber
\delta\rho_\text{r}&(\vc{r},\vc{q}\sigma) =\\&
\sum\limits_{\vc{a}\alpha} \left\{\rho^{0}_{\alpha} \left(\vc{r} -
\left[\vc{R}_{\alpha}^\vc{a}(\vc{q}\sigma) +
\vc{u}_{\alpha}^{\vc{a}}(\vc{q}\sigma)\right]\right) -
\rho_{\alpha}^{0}(\vc{r} - \vc{R}^\vc{a}_{\alpha})\right\}.\label{29}
\end{align}
$\rho_{\alpha}^{0}$ is the density of the unperturbed ion from Eq.
\eqref{1} and $\vc{u}_{\alpha}^\vc{a}$ the displacement of an ion in
Eq. \eqref{19}. In the figures shown, $\delta\rho > 0$ (full lines)
means that electrons are accumulated in the associated region of space.
The broken lines on the other hand indicate that electrons are
depleted. The rigid part of the charge response can clearly be seen in
Fig. \ref{FIG14}a near the displaced  ions, while the nonlocal,
nonrigid part in Fig. \ref{FIG14}b demonstrates that the moving oxygen
O$_{x}$ ions in the half breathing mode generate via nonlocal EPI
changes of the potentials at the silent Cu and O$_{y}$ ions resulting
in corresponding CF's in form of a charge transfer within and between
the CuO$_{y}$ chains. Note, that there are no changes of the transfer
integral between $d$- and $p$-orbitals for the silent Cu and O$_{y}$
ions, nevertheless there is a charge transfer, nonlocally induced. In
case of the \obx mode, Fig. \ref{FIG14}e, the moving O$_{xy}$ ions
induce CF's at the silent Cu ion and we obtain an electronic charge
transfer from that Cu ion where the CuO bonds are compressed to the Cu
ion where the bonds are stretched. According to these calculations we
have a strong coupling of the phonon modes to the charge dynamics. The
nonlocal EPI leads to a (dynamic) charge ordering by CF's in the form
of localized stripes of alternating sign in the CuO plane which are
interacting with the lattice vibrations and with each other. In case of
the \dhalbe anomaly (half breathing mode) the charge stripes are
directed along the $x$- or $y$-axis, respectively, and for \obx (planar
breathing mode) along the diagonals of the CuO plane. The charge
patterns appear instantaneously because of the adiabatic approximation
used in the calculation which is sufficient for these modes. It should
be remarked that qualitatively the same density redistributions are
calculated for \dhalbe and \obx in the metallic phase of LaCuO, see
Refs. \onlinecite{19,21,23,43}.

The nonlocal induced CF's for the \ozz mode in NdCuO is shown in Figs.
\ref{FIG14}c, d. Here we have the situation that the displacement of
the O$_{z}$ ions (Fig. \ref{FIG09} and \ref{FIG14}d) in the ionic
layers excites nonlocally changes of the potential felt by the
electrons in the CuO plane which on her part are responsible for
superconductivity. These facts visualize the importance of such
$c$-axis polarized phonons for superconductivity via the
electron-phonon mechanism.

\begin{figure*}%
\begin{minipage}{0.3\linewidth}%
  \centering%
  \subfigure[\dhalbe]{\includegraphics[width=\linewidth]{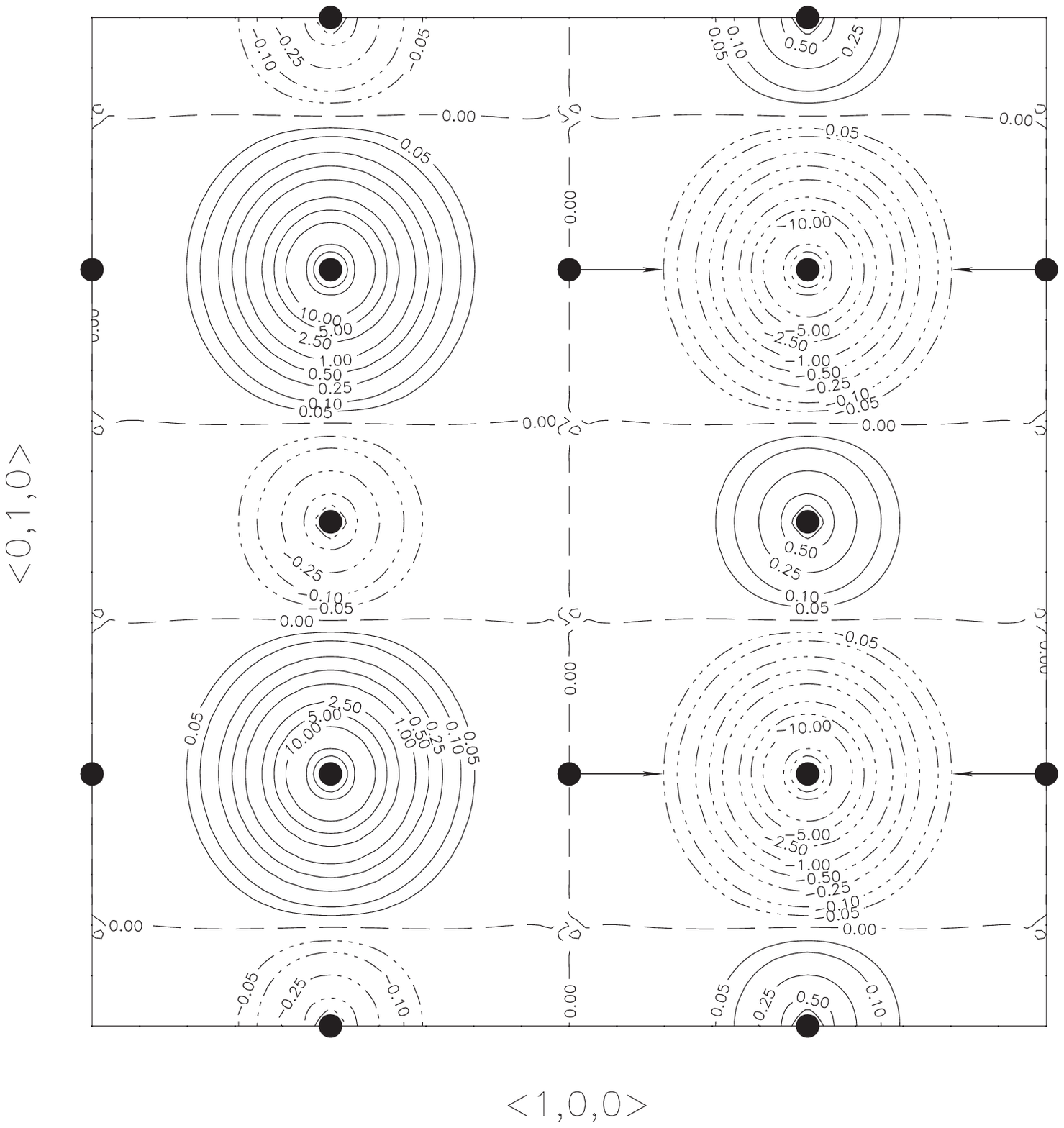}}%
\end{minipage}%
\hspace{0.04\linewidth}%
\begin{minipage}{0.3\linewidth}%
  \centering%
  \subfigure[\ozz]{\includegraphics[width=\linewidth]{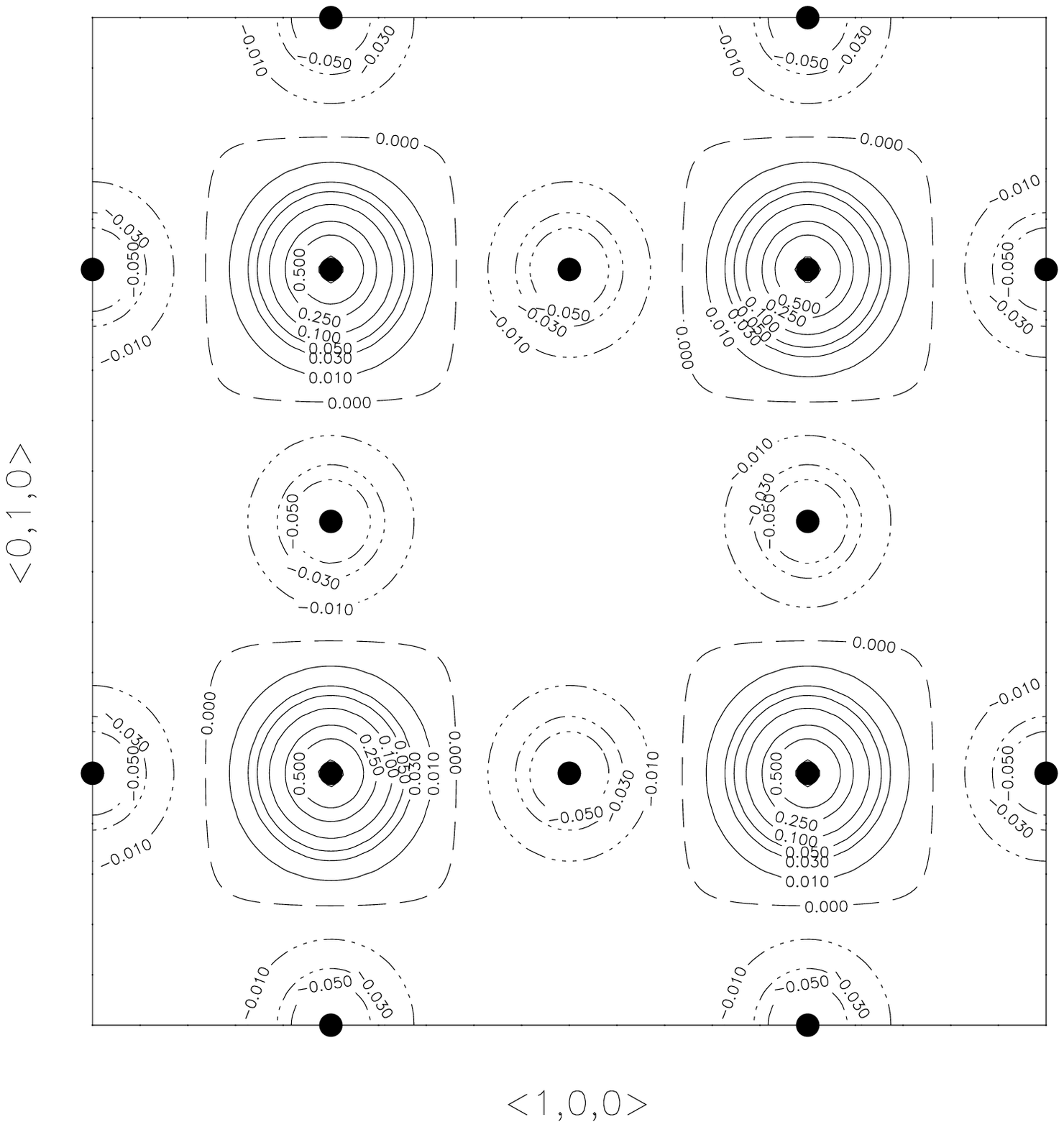}}%
\end{minipage}%
\hspace{0.04\linewidth}%
\begin{minipage}{0.3\linewidth}%
  \centering%
  \subfigure[\obx]{\includegraphics[width=\linewidth]{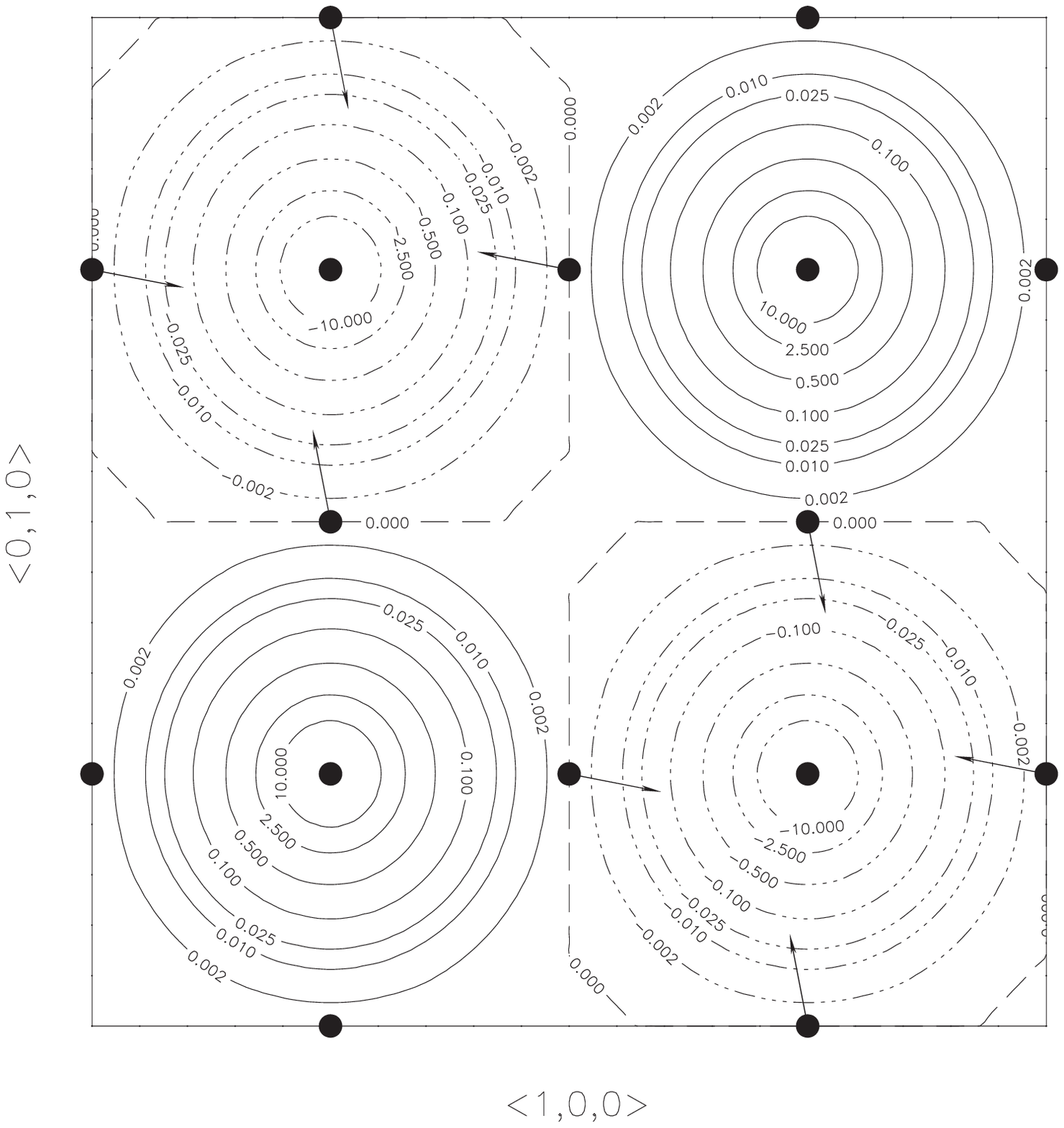}}%
\end{minipage}%
  \caption{Same as in Fig. \ref{FIG14} but for insulating NdCuO. (a)
$\delta\rho_\text{n}$ for the \dhalbe anomaly. (b)
$\delta\rho_\text{n}$ for \ozz. (c) $\delta\rho_\text{n}$ for
\obx.}\label{FIG15}%
\end{figure*}

The nonlocal coupling effects are an expression of the strong component
of the ionic binding along the $c$-axis in the HTSC's, and similar
charge redistributions as in Figs. \ref{FIG14}c, d also have been
calculated for the apex oxygen breathing mode in LaCuO \cite{21}. These
long-ranged Coulomb coupling effects are very special to the HTSC's and
would not be possible in a conventional metal or superconductor because
of local-screening by a high-density electron gas. The strong softening
of about 2.3 THz for the \ozz mode when going from the insulating- to
the metallic phase can physically be understood by comparing the
phonon-induced charge rearrangements for \ozz as calculated in the
metallic- (Figs. \ref{FIG14}c, d) and insulating state (Fig.
\ref{FIG15}b). As shown in Fig. \ref{FIG09} and Fig. \ref{FIG14}d the
O$_{z}$ ions move in phase against the CuO layers, similar as the apex
oxygen mode in LaCuO. Hence, because of the weak screening one can
expect this vibration to induce CF's in the CuO planes as seen.
However, these CF's are \textit{qualitatively} different in the
metallic- and insulating state, respectively, because of the energy gap
for charge excitations in the insulator. This helps to understand the
anomalous softening of the mode across the insulator-metal transition.
As has been shown in Ref. \onlinecite{20} the CF's
$\delta\zeta_{\kappa}(\vc{q}\sigma)$ from Eq. \eqref{18} are
constrained in the insulator with CF's allowed at the Cu and O$_{xy}$
sublattices according to the following sum rule:
\begin{equation}
\sum\limits_{\kappa} \delta\zeta_{\kappa}(\Lambda\sigma) = 0.
\label{30}
\end{equation}
$\kappa$ denotes the CF's in the CuO layer. $\Lambda \sim (0,0,1)$.
Thus, Eq. \eqref{30} particularly holds at the $Z$ point. In contrast
to the constrained expressed by Eq. \eqref{30} for the insulating
state, which means that local charge neutrality of the cell is
maintained under a perturbation due to \ozz, no such a restriction is
present for the metallic state. Consequently, in the insulating state
only \textit{intralayer} charge transfer according to Eq. \eqref{30}
are allowed, see Fig. \ref{FIG15}b. On the other hand, in the metallic
state \ozz induces CF's at Cu and O$_{xy}$ of the same sign in the
whole CuO layer (Fig. \ref{FIG14}c, d). This finally makes an
\textit{interlayer} charge transfer possible, which appears
instantaneously in the adiabatic approximation and provides an
effective screening mechanism for the long-range Coulomb-interaction
and a corresponding softening of \ozz. As already remarked the weak
interlayer coupling in the very anisotropic cuprates most likely makes
necessary a nonadiabatic calculation for modes in a small
$\vc{q}$-space region around the $c$-axis including \ozz. In this case
we find in our calculations for LaCuO a nonadiabatic, insulator-like
charge response crossing over to a coherent adiabatic metallic regime
outside this region \cite{19,38}. Within such a treatment the massive
line broadening of $O_{z}^{Z}$ found experimentally in LaCuO can be
understood. According to Ref. \onlinecite{18} the \ozz mode in metallic
NdCuO is also not well defined indicating a large intrinsic linewidth
which may be understood along the same lines as in LaCuO. Eventually,
the instantaneous interlayer charge transfer in the adiabatic
approximation is replaced by a $c$-axis plasmon at the $Z$ point in the
nonadiabatic regime.

Comparing the results in Fig. \ref{FIG15}a for the charge
redistribution due to the half-breathing mode for insulating NdCuO with
corresponding results of LaCuO we find that the stripe patterns
directed in $x$- or $y$-direction, respectively, found in LaCuO is
destroyed in NdCuO most likely due to the small CF-polarizability of
the O$2p$ orbital in NdCuO. On the other hand, we find for \obx
qualitatively the same diagonal stripe pattern as in LaCuO. On the
whole, the charge redistribution of the OBSM in the insulating state
are more localized around the ions as in the metallic state.

\section{PHONON DENSITY OF STATES}\label{sec_dos}
\begin{figure}
\includegraphics[width=\linewidth]{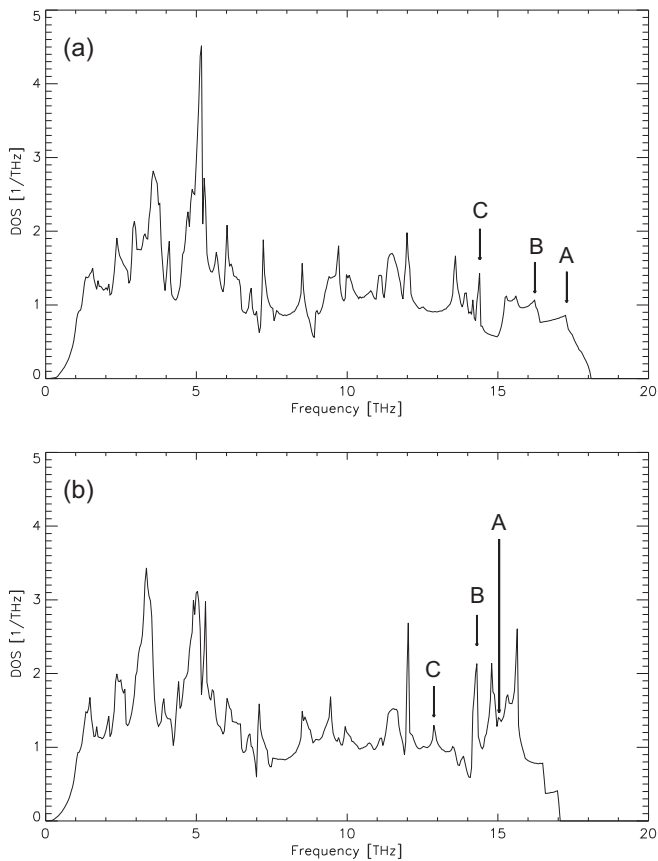}
\caption{Phonon density of states for NdCuO as calculated for the
insulating state (a) and the metallic state (b), respectively. The
characters A, B, C denote the OBSM \ozz, \obx, and the \dhalbe anomaly,
respectively.}\label{FIG16}
\end{figure}

\begin{figure*}
\includegraphics[width=\linewidth]{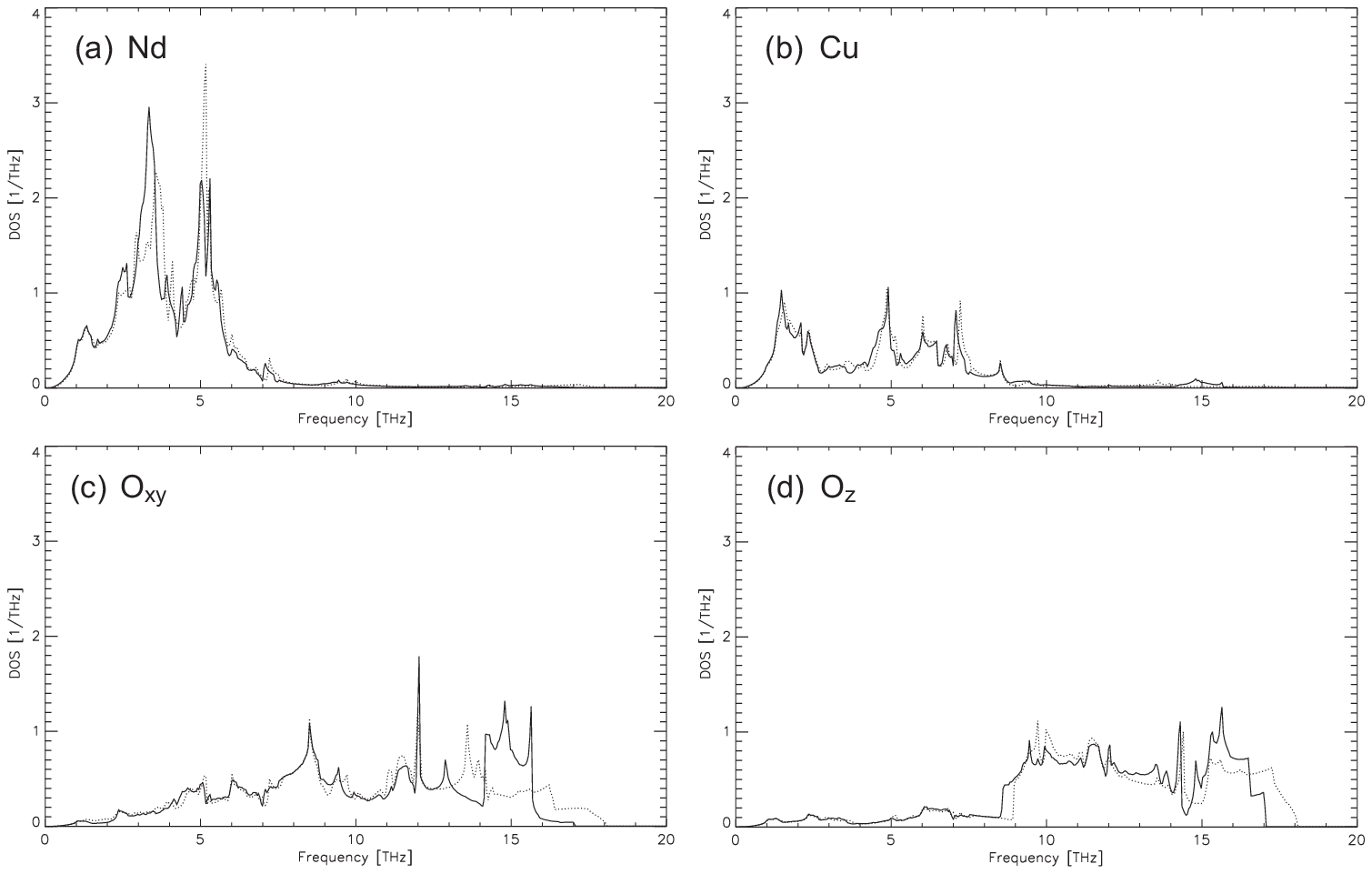}
\caption{Comparison of the atom resolved phonon density of
states of NdCuO between the insulating state ($\cdots$) and the
metallic state ($-\!\!\!-$).}\label{FIG17}
\end{figure*}

In the last topic of this paper we present calculations for the
phonon-density of states of NdCuO in the insulating and metallic state
according to our modeling. Comparing the spectrum of the insulating
state in Fig. \ref{FIG16}a with that of the metallic state in Fig.
\ref{FIG16}b and the calculated results for the corresponding
atom-resolved partial density of states (PDOS) in Fig. \ref{FIG17}
characteristic changes across the insulator-metal transition can be
detected. From the calculated results we find across the transition a
decrease of the width of the spectrum together with a redistribution of
spectral weight in particular in the high-frequency part. The latter
redistribution is essentially due to the softening of the anomalous
OBSM, denoted as A(\ozz), B(\obx) and C(\dhalbe) in Fig. \ref{FIG16}.

Fig. \ref{FIG17}a demonstrates that the low frequency part up to about
6 THz of the spectrum is dominated by the vibrations of the Nd ions.
From the results for the PDOS of the Cu ions in Fig. \ref{FIG17}b,
which are important up to about 8 THz, we conclude that the effect of
the transition is marginal. However, Figs. 17c and 17d display the
essential effect of the transition on the phonon spectrum in the
high-frequency range. Similar as in the case of LaCuO \cite{21} there
is a characteristic shift to lower frequencies of the spectral weight
for the high-frequency oxygen phonon modes across the transition,
indicating again the strong electron-phonon interaction in the metallic
state. Moreover, we do not have just a rigid shift but a characteristic
rearrangement of the high-frequency part of the spectrum related to the
development of the OBSM phonon anomalies in the metallic phase.

The calculated phonon density of states can be compared with the
inelastic-neutron-scattering results as obtained in Ref.
\onlinecite{54} for metallic NdCuO. There are obvious peaks at 3.1,
12.3, and 15.7 THz in the experiments which correlate very well with
significant peaks in the calculated spectrum, shown in Fig.
\ref{FIG16}b. From Fig. \ref{FIG17} we extract, that the peak at 3.1
THz is mainly due to the Nd vibrations with only a small contribution
from the other ions. The peak at 12.3 THz is dominated by the O$_{xy}$
vibrations with a considerable contribution from O$_z$. Finally, in
case of the peak at 15.7 THz, both, O$_{xy}$ and O$_{z}$ participate
with about equal strength.

\section{SUMMARY AND CONCLUSIONS}\label{sec_summary}
We have calculated for NdCuO complete phonon dispersion curves, total-
and atom-resolved density of states and local as well as nonlocal
phonon-induced charge redistributions across the insulator-metal
transition. We have used our microscopic modeling for the undoped
insulating-, the underdoped strange metallic- and the optimally doped
metallic state in terms of orbital selective
incompressibility-compressibility transitions, different from the
$p$-doped case studied earlier. In particular, we have investigated the
possibly generic phonon anomalies (OBSM) in the HTSC's and their
complex anticrossing behaviour in NdCuO. A good agreement with the
corresponding experimental phonon dispersion measured by INS has been
found. Our calculations support the assignments of the anomalies as
provided by INS as compared to the INX results.

In order to obtain the calculated results, both, long-ranged Coulomb
interactions of ionic origin as well as short-ranged repulsive
interactions of electronic origin together with a sufficiently large
set of orbital degrees of freedom are necessary. We conclude that a
purely ionic model, similar as in case of LaCuO, leads to a
considerable overestimation of the width of the phonon spectrum and
some unstable branches. Including covalent corrections a suitable
reference model for the insulating phase of the HTSC's can be
constructed that cures the disadvantages of the purely ionic
description. Starting from such a rigid reference model it has been
shown that nonlocal, nonrigid electronic polarization processes via
localized CF's and DF's are crucial for a correct specification of the
density response and phonon dynamics.

From our calculations some insight into the strange metallic state of
the cuprates and an electron-hole asymmetry introduced by doping can be
gained. We have found a characteristically different behaviour of the
O$2p$ orbital in the charge response when comparing $p$-doped LaCuO to
$n$-doped NdCuO, respectively. In both, the insulating and metallic
phase, the electronic CF-polarizability of O$2p$ is strongly supressed
in NdCuO as compared with LaCuO, indicating a stronger localization of
the O$2p$ orbitals in the $n$-doped material. This points to an
enhancement of the ionic component of binding. Simultaneously, the
stronger localization of O$2p$ tends to stabilize the antiferromagnetic
spin correlations in comparison to a $p$-doped material like LaCuO,
because a localization of the O$2p$ orbital can be expexted to be
favorable for superexchange. Note, that for the $n$-type cuprates the
antiferromagnetic phase extends much further with doping and from our
calculations one might speculate that the polarizability of oxygen
should play an indirect role for antiferromagnetism and
superconductivity.

In our modeling we found for NdCuO an increased matrix element for the
Cu$4s$ CF-polarizability in the metallic state as compared to LaCuO. An
increased occupation of this delocalized orbital in the hybridized
metallic state at the Fermi energy, likewise as in LaCuO is very
important to comprehend the softening behaviour of the OBSM, \obx-,
\ozz-, \dhalbe-anomaly, found across the insulator-metal transition via
the underdoped state. The enhanced Cu$4s$-polarizability in a $n$-doped
material seems to be consistent with an electron-hole asymmetry
introduced by the doping process.

We have presented a qualitative physical picture of the electronic
state in the cuprates consistent with our modeling of the charge
response which on his part leads to a good agreement of the calculated
phonon dispersion with the INS results, in particular for the
\textit{generic} phonon anomalies. These modes along with $c$-axis
modes like \ozz from the nonadiabatic region most probably should
contribute to electron self-energy corrections like kinks and
pseudogaps seen in ARPES experiments as well as to the unusual
transport properties along the $c$-axis in the cuprates. The definite
calculations have shown that the full set of orbital degrees of
freedom, i.e. Cu$3d$/$4s$ and O$2p$ is essential. Qualitatively $s$-
and $d$-wave superconductivity has been discussed within our modeling
and it is conjectured that holes in the O$2p$ states should play a
similar role for $d$-wave superconductivity in \textit{both}, $p$-type
as well as $n$-type cuprate superconductors.

From our calculation of the complex anticrossing scenario in NdCuO seen
in the experiments but being absent in LaCuO we conclude that the
anticrossing is related to the different crystal structure of these
materials.

Our calculations of the phonon-induced charge redistributions performed
for the anomalous OBSM, \obx, \dhalbe, indicate that the nonlocal EPI
of ionic origin leads to (dynamic) charge ordering by CF's in the form
of localized stripes of alternating sign in the CuO plane, minimizing
the energy of the system and decreasing the phonon frequencies
considerably.

The strong frequency renormalization of \ozz across the insulator-metal
transition is attributed by our calculation to an \textit{interlayer}
charge transfer in the metallic state not possible in the insulator
where only \textit{intralayer} charge transfer is allowed.

Finally, there is, similar as in LaCuO, a characteristic shift to lower
frequencies of the spectral weight for the high frequency oxygen
vibrations in the metallic phase accompanied by a strong rearrangement
of the high-frequency part of the phonon density of states. This is
essentially related to the development of the anomalous OBSM in the
metallic phase, due to a strong electron-phonon interaction, also in
the $n$-doped compounds.

\section*{Acknowledgments}
We thank L. Pintschovius for the careful reading of the manuscript.

\end{document}